\documentclass[12pt]{article}
\usepackage[symbol]{footmisc}
\def\correspondingauthor{\footnote{Corresponding author.  }}
\usepackage[left=2cm,top=2.5cm,right=2cm,bottom=2.5cm]{geometry}
\usepackage{amsmath, amssymb}

\usepackage{graphicx}
\usepackage{graphics}
\usepackage{epstopdf}
\usepackage{subfigure}
\usepackage{amsfonts}
\usepackage{sectsty}
\usepackage{sectsty}
\usepackage{hyperref}
\usepackage{cite}

\begin{document}
	
	\begin{center}
	\large{\bf{Traversable Wormholes and Energy Conditions with Two Different Shape Functions in $f(R)$ Gravity}} \\
	\vspace{5mm}
			\normalsize{Nisha Godani$^1$ and Gauranga C. Samanta$^{2,}{}$\correspondingauthor{}, }\\
						\normalsize{$^1$ Department of Mathematics, Institute of Applied Sciences and Humanities\\ GLA University, Mathura, Uttar Pradesh, India\\
				$^2$ Department of Mathematics, BITS Pilani K K Birla Goa Campus, Goa, India}\\
				\normalsize {nishagodani.dei@gmail.com\\gauranga81@gmail.com}
\end{center}
	
	\begin{abstract}
	Traversable wormholes, tunnel like structures introduced by Morris \& Thorne \cite{morris1}, have a significant role in connection of two different space-times or two different parts of the same space-time. The characteristics of these wormholes depend upon the redshift and shape functions which are defined in terms of radial coordinate. In literature, several shape functions are defined and wormholes are studied in $f(R)$ gravity  with respect to these shape functions \cite{lobo, saiedi, baha}.  In this paper, two shape functions (i) $b(r)=\dfrac{{r_0} \log (r+1)}{\log ({r_0}+1)}$  and (ii) $b(r)=r_0(\frac{r}{r_0})^\gamma$, $0<\gamma<1$ are considered. The first shape function is newly defined, however the second one is collected from the literature\cite{cataldo}.
	The wormholes are investigated for each type of shape function  in $f(R)$ gravity with $f(R)=R+\alpha R^m-\beta R^{-n}$, where $m$, $n$, $\alpha$, and $\beta$ are real constants. Varying parameters $\alpha$ or $\beta$,  $f(R)$ model is studied in five subcases for each type of shape function. In each case, the energy density, radial \& tangential pressures, energy conditions that include null energy condition, weak energy condition, strong energy condition \& dominated energy condition, and anisotropic parameter are computed.   The energy density is found to be positive and all energy conditions are obtained to be violated which supports the existence of wormholes. Also, the equation of state parameter is obtained to possess values less than -1, that shows the presence of the phantom fluid and leads towards the expansion of the universe.
\end{abstract}

\section{Introduction}
A wormhole has a tube-like structure which is asymptotically flat on both sides. It  gives rise to the geometries that connect any two points of the same space-time or two different space-times.  A wormhole is static or non-static in accordance with the constant or variable radius of its throat. Wormholes have been studied in various aspects in literature \cite{morris1, haw, fried, fro, visshoch}

These arise as a non-vacuum solution of Einstein's field equations. Flamm \cite{flamm} first studied wormhole type solutions in Einstein gravity, but his solutions were found to be unstable. Then Einstein and Rosen \cite{eins-ros}  represented particle by a bridge which is known as Einstein-Rosen bridge and carried out a comprehensive study of wormhole solutions. Wheeler \cite{wheel} obtained Kerr wormholes which are the objects of quantum foam, connects different parts of space-time and operates at the Planck scale. But these wormholes were not traversable. Thorne and his student Morris \cite{morris1}, understood the structure of  wormholes with a throat \& two mouths
and introduced static traversable wormholes and sparked this area. They used the principles of general relativity and found a possible way of time travelling. The existence of wormholes demands the presence of some exotic matter that violates the energy conditions \cite{morris2, visser}. But there is a lack of the matter bearing the geometry of the wormhole. This led to the investigation that whether the modified theory of gravity explaining the acceleration of the universe can describe the wormhole structure.

In the literature, several theories have been introduced and studied. Among these, $f(R)$ theory of gravity is found to be a suitable candidate theory which modifies Einstein's theory of general relativity by replacing the gravitational action $R$ by an arbitrary function $f(R)$, where $R$ is the Ricci scalar. This theory provides more complex equations than general relativity and gives a larger set of solutions. Using this theory, Starobinsky \cite{star} first proposed an accelerated model of inflation. Sotiriou \cite{sotiriou} studied the conditions for the equivalence of $f(R)$ and scalar tensor theories. Nojiri and Odintsov \cite{Nojiri} investigated a number of modified theories and  discussed their properties.  They provided cosmological reconstruction of modified theories and discussed Big Rip \& future singularities. They also investigated various realistic $f(R)$ models that unify the inflation with dark energy.
Huang \cite{huang} investigated an $f(R)$ model of inflation. Many other cosmologists have also investigated $f(R)$ theory of gravity in various aspects
\cite{noj, cog2008, felice, Bamba4, Bamba3, Bamba1, sebas, thakur, Bamba, Peng, Motohashi, Astashenok, noj2017, baha1, Yousaf, Faraoni, Abbas, Sussman, Mongwane, Mansour, Muller, Wang, Papagiannopoulos, Oikonomou, Chakraborty, Nashed, Capozziello, Gu, Abbas1, Mishra, Odintsov}.

Numerous efforts have also been put for the exploration of wormhole geometries. The wormholes violates the energy conditions, therefore they do not appear as a solution of classical gravity with matter and indicates the possibility of occurrence  of wormholes at quantum level. Hochberg et al. \cite{Hochberg} obtained solution of semi classical  Einstein field equations representing wormholes. Nojiri et al. \cite{Nojiri1} obtained the possibility of inducing wormholes in early time using effective equation method. This approach is limited only with the scalar matter. Successively, Nojiri et al. \cite{Nojiri2} answered about the existence of spherically symmetric wormholes from GUTs at the early universe. They used large N, 4d anomaly induced one loop effective action and some initial conditions and proved that the wormhole structures can be produced numerically due to GUTs effects at early times.
Lemos et al. \cite{lemos} reviewed the ideas developed in the field of wormholes, analyzed the traversable wormholes in the occurrence of generic cosmological constant and studied several physical properties due to the presence of cosmological constant term.
Furey \& De Bendictis \cite{furey} adopted non-linear powers of Ricci scalar in gravitational action and studied wormholes. They obtained the existence of static wormhole satisfying the weak energy condition.
Dotti et al. \cite{dotti} considered higher dimensional gravity and studied wormhole solutions in vacuum.
Spherically symmetric traversable wormholes have also been studied using modified Chaplygin gas in \cite{lobo1, chakra, jamil}.
Lobo and Oliveira \cite{lobo} developed traversable wormhole structures in modified $f(R)$ gravity. They found the higher order curvature derivative terms responsible for the violation of the null energy condition and supporting the nonstandard wormhole structures. They considered constant redshift function, some specific shape functions and various equations of state to find the exact solutions.
Cataldo et al. \cite{cata} presented Lorentzian wormhole solutions for Einstien's field equations. They used barotropic equation of state for radial and lateral pressures and explored static and evolving wormholes in $N + 1$ dimensions.
Saiedi and Esfahani \cite{saiedi} considered shape and redshift functions are constant and scale factor as some positive power of cosmic time. They constructed wormhole solutions in $f(R)$ gravity and inspected null and weak energy conditions.
Bouhmadi-L\'{o}pez et al. \cite{lopez} considered the sum of energy density and radial pressure to be proportional to a constant less than the area of the wormhole mouth. They examined the solutions of spherically symmetric wormhole and analyze the stability regions.
Najafi et al. \cite{najafi} took an extra space-like dimension and studied traversable wormhole in FLRW model. They analyzed the effect of extra dimension on energy density, scale factor and shape function.
Bahamonde et al. \cite{baha} studied cosmological wormhole in $f(R)$ theory of gravity. They built a dynamical wormhole asymptotically approaching towards the FLRW universe and used the approximation of small wormholes for analysis. For the wormholes they considered, it was found that the presence of exotic matter near the throat is not needed, however it is always needed in case of general relativity.
Rahaman et al. \cite{Rahaman} studied wormhole solutions in Finslerian structure of space-time. They presented a wide variety of solutions and explored wormhole geometry by considering different choices of shape function and energy density.
Zubair et al. \cite{zubair} investigated wormhole solutions in the context of generalized $f(R, \phi)$ gravity for three types of fluids. They explored energy conditions and obtained wormhole solutions without need of exotic matter.
Kuhfittig \cite{peter} considered non-commutative geometry and discussed the existence of wormholes in $f(R)$ gravity. He considered various shape functions and obtained wormhole solutions satisfying general properties. He also considered $f(R) = \alpha R^2$ and determined wormhole solutions.
Novikov \cite{novi} reviewed wormholes and categorize them into three classes. They determined the properties of wormholes and described the  relation between black holes and wormholes.
Subsequently, many authors have been studied wormholes in different contexts \cite{Zangeneh1, Mehdizadeha, Zangeneh, Moraes, Bejarano, Mehdizadeh, Rogatko, Paul, Shaikh, Ovgun, Tsukamoto}.
Recently, Barros and Lobo \cite{barros} used three form fields and studied  static and spherically symmetric wormhole structures. They found various numerical and analytical solutions and showed that in the presence of three-form fields the null and weak energy conditions are satisfied in whole space-time.

The organization of the paper is as follows. In Sec-2, the field equations are obtained for wormhole geometry in $f(R)$ gravity.
In Sec-3, wormhole solutions in $f(R)$ model with two different  shape functions are obtained. Results and
conclusions are finally provided in Sec-4.

\section{Wormhole Geometry in $f(R)$ Gravity}
The $f(R)$  theory of gravity is  a generalization of Einstein's theory of relativity because it replaces the  gravitational action $R$ of general relativity by a general function $f(R)$ of $R$. The gravitational action for this theory  is defined as
\begin{equation}\label{action}
S_G=\dfrac{1}{16\pi}\int[f(R) + L_m]\sqrt{-g}d^4x,
\end{equation}
where $L_m$ is the matter Lagrangian density and  $g$ is the  determinant of the metric $g_{\mu\nu}$. Varying Eq.(\ref{action}) with respect to the metric $g_{\mu\nu}$, the field equations are
\begin{equation}
f'(R)R_{\mu\nu} -\dfrac{1}{2}f(R)g_{\mu\nu}-\triangledown_\mu\triangledown_\nu f'(R)+\square f'(R)g_{\mu\nu}= T_{\mu\nu},
\end{equation}	
where $R_{\mu\nu}$ is the Ricci tensor, $f^{'}(R)=\frac{df}{dR}$, $R$ is the curvature scalar and $T_{\mu\nu}=\frac{\partial L_m}{\partial g^{\mu\nu}}$ is the energy momentum tensor for the matter source of the wormholes, it is  defined as
\begin{equation}
T_{\mu\nu} = (\rho + p_t)u_\mu u_\nu - p_tg_{\mu\nu}+(p_r-p_t)X_\mu X_\nu,
\end{equation}	
such that
\begin{equation}
u^{\mu}u_\mu=-1 \mbox{ and } X^{\mu}X_\mu=1,
\end{equation}
where $\rho$,  $p_t$ and $p_r$, respectively, stand for the energy density, tangential pressure and radial pressure.
The static, spherical and symmetric metric defining the geometry of wormhole is  \begin{equation}
ds^2=-e^{2\Phi(r)}dt^2+\frac{dr^2}{1-b(r)/r} + r^2(d\theta^2+\sin^2\theta^2\Phi^2),
\end{equation}
where $r$ is the radial coordinate, $\Phi(r)$ and $b(r)$ are arbitrary functions of radius $r$ only.  Where $b(r)$ decides the spatial shape of the wormhole, hence we can say it the ``shape function'' and $\Phi(r)$ determines the gravitational redshift, hence we can say it the ``redshift function''.
The variable $r$ takes values from $r_0$ to infinity, where $r_0$ is the radius of the throat of the wormhole.
For the existence of wormhole solutions the shape function should satisfy the following conditions:

\begin{itemize}
  \item [(i)] $b(r_0)=r_0$
  \item [(ii)] $\frac{b(r)-b'(r)r}{b^2}>0$
  \item [(iii)] $b'(r_0)-1\leq 0$
  \item [(iv)] $\frac{b(r)}{r}<1$ for $r>r_0$
  \item [(v)] $\frac{b(r)}{r}\rightarrow 0$ as $r\rightarrow\infty.$
\end{itemize}
For the absence of horizons and singularities, the redshift function should be positive  for all $r>r_0$. In this study, the redshift function $\Phi(r)$ is taken as constant which implies that $\Phi'(r)=0$. The condition  $\Phi'(r)=0$ is known as tidal force solution which is a required property for a traversable wormhole.

The Einstein's field equations in the framework of $f(R)$ gravity take the following form:
\begin{equation}\label{6}
  \rho=F(r)\frac{b'(r)}{r^2}
\end{equation}
\begin{equation}\label{7}
  p_r=-F(R)\frac{b(r)}{r^3}+F'(r)\frac{rb'(r)-b(r)}{2r^2}-F''(r)(1-\frac{b(r)}{r})
\end{equation}
\begin{equation}\label{8}
  p_t=-\frac{F'(r)}{r}(1-\frac{b(r)}{r})+\frac{F(r)}{2r^3}(b(r)-rb'(r)),
\end{equation}
where $F$ denotes the derivative of $f(R)$ with respect to Ricci scalar $R$ given by
$R(r)=\frac{2b'(r)}{r^2}$ and prime upon a function denotes the derivative of that function with respect to  radial coordinate $r$. The equation of state parameter in terms of radial pressure is called radial state parameter and is defined as
\begin{equation}
w = \frac{p_r}{\rho}.
\end{equation}
The mass function is defined as
\begin{equation} \label{mass}
m=\int_{r_0}^{r}4\pi r^2\rho dr.
\end{equation}
Further, the anisotropy parameter in terms of radial and tangential pressures is defined as
\begin{equation}
\triangle=p_t-p_r.
\end{equation}
The attractive and repulsive nature of the geometry depends on the value of the anisotropy parameter $\triangle$.
If $\triangle$ is negative, then geometry is said to attractive, if it is positive, then geometry is said to be repulsive and if it is zero, then the geometry contains isotropic pressure.

\section{Wormholes in $f(R)$ Model with Different Shape Functions}
In literature, several $f(R)$ models have been proposed and investigated \cite{star, cataldo, noj1, storiou6, meng4, hu, Dutta, Sussman, Oikonomou, Sharif}.
However, in this paper the model $f(R)=R-\frac{\alpha}{(R-\lambda_1)^n}+\beta (R-\lambda_2)^m$ proposed by Nojiri and Odintsov \cite{noj1} is considered.
By introducing the auxiliary fields $U$ and $V$, one may rewrite the action as follows:
\begin{equation}\label{odaction}
  S=\frac{1}{\kappa^2}\int \sqrt{-g}[V(R-U)+f(U)]d^4x.
\end{equation}
If we take the variation with respect to $U$, then we can have $V=f'(U)$, which may be solved with respect to $U$ as $U=g(V)$.
Eliminating $U$ in \eqref{odaction} with the help of $U=g(V)$, we get
\begin{equation}\label{action1}
  S=\frac{1}{\kappa^2}\int \sqrt{-g}[V(R-g(V))+f(g(V))]d^4x.
\end{equation}
Also, one may eliminate $V$ by using $V=f'(U)$ and we can have
\begin{equation}\label{action2}
  S=\frac{1}{\kappa^2}\int \sqrt{-g}[f'(U)(R-U)+f(U)].
\end{equation}
The above two equations \eqref{action1} and \eqref{action2} are equivalent, at least classically. By assuming conformal transformation
$g^{\mu\nu}\to e^{\sigma}g_{\mu\nu}$, $d$-dimensional scalar curvature is transformed as
\begin{equation}\label{}
  R^{(d)}\to e^{-\sigma}\left(R^{(d)}-(d-1)\square\sigma-\frac{(d-1)(d-2)}{4}g^{\mu\nu}\partial_{\mu}\sigma\partial_{\nu}\sigma\right).
\end{equation}
For $d=4$, by choosing $\sigma=-\ln f'(U)$, now the action \eqref{action2} can be rewritten as
\begin{equation}\label{}
  S_{E}=\frac{1}{k^2}\int \sqrt{-g}\bigg[R-\frac{3}{2}\left(\frac{f''(U)}{f'(U)}\right)^2g^{\rho\sigma}\partial_{\rho}U\partial_{\sigma}U
  -\frac{Uf'(U)-f(U)}{f'(U)^2}\bigg]d^4x.
\end{equation}
If we use $\sigma=-\ln (f'(U))=-\ln V$, the Einstein action becomes
\begin{equation}\label{}
  S_{E}=\frac{1}{\kappa^2}\int \sqrt{-g}\bigg[R-\frac{3}{2}g^{\rho\sigma}\partial_{\rho}\sigma\partial_{\sigma}\sigma -V_1(\sigma)\bigg]d^4x,
\end{equation}
where $V_1(\sigma)=e^{\sigma}g(e^{-\sigma})-e^{2\sigma} f(g(e^{-\sigma}))=\frac{Uf'(U)-f(U)}{f'(U)^2}$.
it is assumed $f'(U)>0$. If $f'(A)<0$, then we may define $\sigma=-\ln |f'(U)|$. So, the sign in front of the scalar curvature
becomes negative. In other words, antigravity could be generated. As a specific choice of $f(R)$,
Nojiri and Odintsov \cite{noj1} considered
\begin{equation}\label{}
  f(R)=R-\frac{\alpha}{(R-\lambda_1)^n}+\beta (R-\lambda_2)^m,
\end{equation}
where the coefficients $n, m, \alpha, \beta $ are real constants.

In this paper, we are motivated by the work of Nojiri and Odintsov \cite{noj1} and considered the $f(R)$ model in the form

\begin{equation}
f(R) = R + \alpha R^m - \beta R^{-n},
\end{equation} where $m$, $n$, $\alpha$ and $\beta$ are real constants. For large curvature, the term $R^m$ dominate and produce inflation at early times, however for low curvature the term $R^{-n}$ dominate and produce current cosmic acceleration. Recently, Cao et al. \cite{cao} used this form of $f(R)$ to study FRW model with $f(R)$ gravity in Palatini formalism and found it to lead towards the late time acceleration. In the present paper, this form of $f(R)$ is considered with the following two shape functions  to study the wormhole solutions:
\begin{itemize}
  \item [(i)] $b(r)=\dfrac{{r_0} \log (r+1)}{\log ({r_0}+1)},$
  \item [(ii)] $b(r)=r_0(\frac{r}{r_0})^\gamma$, $0<\gamma<1.$
\end{itemize}
 These shape functions should satisfy the required conditions discussed in Section 2 for the existence of wormhole solutions.

\subsection{Case I:}
In this case, the wormhole solutions are obtained by using the shape function $b(r)=\dfrac{{r_0} \log (r+1)}{\log ({r_0}+1)}$. The energy density, radial pressure and tangential pressure  are obtained from the field equations \eqref{6} - \eqref{8} as follows:


\begin{eqnarray} \label{11}
\rho&=&
\frac{{r_0} \left(2^{-n-1} \left(\frac{{r_0}}{r^2 (r+1) \log ({r_0}+1)}\right)^{-n-1} \left(\alpha  m 2^{m+n} \left(\frac{{r_0}}{r^2 (r+1) \log ({r_0}+1)}\right)^{m+n}+\beta  n\right)+1\right)}{r^2 (r+1) \log ({r_0}+1)}
\end{eqnarray}

\begin{eqnarray}\label{12}
p_r&=&2^{-n-3} ((r+1) \log (r+1)-r) \left(\frac{{r_0}}{r^2 (r+1) \log ({r_0}+1)}\right)^{-n-1}\nonumber\\
 &\times& \bigg[\alpha  (m-1) m \left(-2^{m+n}\right) \left(\frac{{r_0}}{r^2 (r+1) \log ({r_0}+1)}\right)^{m+n}
+\beta  n^2+\beta  n\bigg] \nonumber\\
&-&\frac{1}{r^3 \log ({r_0}+1)}{r_0}\log (r+1)
\times\bigg[2^{-n-1} \left(\alpha  m 2^{m+n} \left(\frac{{r_0}}{r^2 (r+1) \log ({r_0}+1)}\right)^{m+n}+\beta  n\right) \nonumber\\
&+&1
\left(\frac{{r_0}}{r^2 (r+1) \log ({r_0}+1)}\right)^{-n-1}\bigg]
-\frac{1}{{r_0}^3}2^{-n-3}r^5(r+1)^3 \nonumber\\
&\times&\left(m \left(m^2-3 m+2\right)\alpha 2^{m+n}+\beta n^3+3 \beta  n^2+2 \beta  n
\left(\frac{r_0}{r^2 (r+1) \log ({r_0}+1)}\right)^{m+n}\right)\nonumber\\
&\times&\left(\frac{r_0}{r^2 (r+1) \log ({r_0}+1)}\right)^{-n}\log ^2({r_0}+1)r \log ({r_0}+1)-{r_0} \log (r+1)
\end{eqnarray}

\begin{eqnarray}\label{13}
p_t&=&-\frac{1}{2 r^3 (r+1) \log ({r_0}+1)}\bigg[{r_0}r-(r+1) \log (r+1)\bigg[(1+2^{-n-1} \bigg[(\alpha  m 2^{m+n} \nonumber\\
 &\times&\left(\frac{{r_0}}{r^2 (r+1) \log ({r_0}+1)}\right)^{m+n}+\beta  n\bigg]\left(\frac{{r_0}}{r^2 (r+1) \log ({r_0}+1)}\right)^{-n-1}\bigg]\nonumber\\
 &+&2^{-n-1} r (r+1) \bigg[\alpha  (m-1) m \left(-2^{m+n}\right) \left(\frac{{r_0}}{r^2 (r+1) \log ({r_0}+1)}\right)^{m+n}+\beta  n^2 \nonumber\\
 &+&\beta  n\bigg]({r_0} \log (r+1)-r \log ({r_0}+1)) \left(\frac{{r_0}}{r^2 (r+1) \log ({r_0}+1)}\right)^{-n-2}\bigg]
\end{eqnarray}

Adding equations \eqref{11} and \eqref{12}, we have

\begin{eqnarray}
\rho+p_r&=&
\frac{{r_0} \left(2^{-n-1} \left(\frac{{r_0}}{r^2 (r+1) \log ({r_0}+1)}\right)^{-n-1} \left(\alpha  m 2^{m+n} \left(\frac{{r_0}}{r^2 (r+1) \log ({r_0}+1)}\right)^{m+n}+\beta  n\right)+1\right)}{r^2 (r+1) \log ({r_0}+1)}\nonumber\\
&+&2^{-n-3} ((r+1) \log (r+1)-r) \left(\frac{{r_0}}{r^2 (r+1) \log ({r_0}+1)}\right)^{-n-1}\nonumber\\
&\times& \bigg[\alpha  (m-1) m \left(-2^{m+n}\right) \left(\frac{{r_0}}{r^2 (r+1) \log ({r_0}+1)}\right)^{m+n}
+\beta  n^2+\beta  n\bigg] \nonumber\\
&-&\frac{1}{r^3 \log ({r_0}+1)}{r_0}\log (r+1)
\times\bigg[2^{-n-1} \left(\alpha  m 2^{m+n} \left(\frac{{r_0}}{r^2 (r+1) \log ({r_0}+1)}\right)^{m+n}+\beta  n\right) \nonumber\\
&+&1
\left(\frac{{r_0}}{r^2 (r+1) \log ({r_0}+1)}\right)^{-n-1}\bigg]
-\frac{1}{{r_0}^3}2^{-n-3}r^5(r+1)^3 \nonumber\\
&\times&\left(m \left(m^2-3 m+2\right)\alpha 2^{m+n}+\beta n^3+3 \beta  n^2+2 \beta  n
\left(\frac{r_0}{r^2 (r+1) \log ({r_0}+1)}\right)^{m+n}\right)\nonumber\\
&\times&\left(\frac{r_0}{r^2 (r+1) \log ({r_0}+1)}\right)^{-n}\log ^2({r_0}+1)r \log ({r_0}+1)-{r_0} \log (r+1)
\end{eqnarray}
Adding equations \eqref{11} and \eqref{13}, we have
\begin{eqnarray}
\rho+p_t&=&
\frac{{r_0} \left(2^{-n-1} \left(\frac{{r_0}}{r^2 (r+1) \log ({r_0}+1)}\right)^{-n-1} \left(\alpha  m 2^{m+n} \left(\frac{{r_0}}{r^2 (r+1) \log ({r_0}+1)}\right)^{m+n}+\beta  n\right)+1\right)}{r^2 (r+1) \log ({r_0}+1)}\nonumber\\
&-&\frac{1}{2 r^3 (r+1) \log ({r_0}+1)}\bigg[{r_0}r-(r+1) \log (r+1)\bigg[(1+2^{-n-1} \bigg[(\alpha  m 2^{m+n} \nonumber\\
&\times&\left(\frac{{r_0}}{r^2 (r+1) \log ({r_0}+1)}\right)^{m+n}+\beta  n\bigg]\left(\frac{{r_0}}{r^2 (r+1) \log ({r_0}+1)}\right)^{-n-1}\bigg]\nonumber\\
&+&2^{-n-1} r (r+1) \bigg[\alpha  (m-1) m \left(-2^{m+n}\right) \left(\frac{{r_0}}{r^2 (r+1) \log ({r_0}+1)}\right)^{m+n}+\beta  n^2 \nonumber\\
&+&\beta  n\bigg]({r_0} \log (r+1)-r \log ({r_0}+1)) \left(\frac{{r_0}}{r^2 (r+1) \log ({r_0}+1)}\right)^{-n-2}\bigg]
\end{eqnarray}
Adding equations \eqref{11}, \eqref{12} and $2\times\eqref{13}$, we have
\begin{eqnarray}
\rho+p_r+2p_t&=&
\frac{{r_0} \left(2^{-n-1} \left(\frac{{r_0}}{r^2 (r+1) \log ({r_0}+1)}\right)^{-n-1} \left(\alpha  m 2^{m+n} \left(\frac{{r_0}}{r^2 (r+1) \log ({r_0}+1)}\right)^{m+n}+\beta  n\right)+1\right)}{r^2 (r+1) \log ({r_0}+1)}\nonumber\\
&+&2^{-n-3} ((r+1) \log (r+1)-r) \left(\frac{{r_0}}{r^2 (r+1) \log ({r_0}+1)}\right)^{-n-1}\nonumber\\
&\times& \bigg[\alpha  (m-1) m \left(-2^{m+n}\right) \left(\frac{{r_0}}{r^2 (r+1) \log ({r_0}+1)}\right)^{m+n}
+\beta  n^2+\beta  n\bigg] \nonumber\\
&-&\frac{1}{r^3 \log ({r_0}+1)}{r_0}\log (r+1)
\times\bigg[2^{-n-1} \left(\alpha  m 2^{m+n} \left(\frac{{r_0}}{r^2 (r+1) \log ({r_0}+1)}\right)^{m+n}+\beta  n\right) \nonumber\\
&+&1
\left(\frac{{r_0}}{r^2 (r+1) \log ({r_0}+1)}\right)^{-n-1}\bigg]
-\frac{1}{{r_0}^3}2^{-n-3}r^5(r+1)^3 \nonumber\\
&\times&\left(m \left(m^2-3 m+2\right)\alpha 2^{m+n}+\beta n^3+3 \beta  n^2+2 \beta  n
\left(\frac{r_0}{r^2 (r+1) \log ({r_0}+1)}\right)^{m+n}\right)\nonumber\\
&\times&\left(\frac{r_0}{r^2 (r+1) \log ({r_0}+1)}\right)^{-n}\log ^2({r_0}+1)r \log ({r_0}+1)-{r_0} \log (r+1)\nonumber\\
&-&\frac{2}{2 r^3 (r+1) \log ({r_0}+1)}\bigg[{r_0}r-(r+1) \log (r+1)\bigg[(1+2^{-n-1} \bigg[(\alpha  m 2^{m+n} \nonumber\\
&\times&\left(\frac{{r_0}}{r^2 (r+1) \log ({r_0}+1)}\right)^{m+n}+\beta  n\bigg]\left(\frac{{r_0}}{r^2 (r+1) \log ({r_0}+1)}\right)^{-n-1}\bigg]\nonumber\\
&+&2^{-n-1} r (r+1) \bigg[\alpha  (m-1) m \left(-2^{m+n}\right) \left(\frac{{r_0}}{r^2 (r+1) \log ({r_0}+1)}\right)^{m+n}+\beta  n^2 \nonumber\\
&+&\beta  n\bigg]({r_0} \log (r+1)-r \log ({r_0}+1)) \left(\frac{{r_0}}{r^2 (r+1) \log ({r_0}+1)}\right)^{-n-2}\bigg]
\end{eqnarray}

\begin{eqnarray}
\rho-\lvert p_r \rvert&=&
\frac{{r_0} \left(2^{-n-1} \left(\frac{{r_0}}{r^2 (r+1) \log ({r_0}+1)}\right)^{-n-1} \left(\alpha  m 2^{m+n} \left(\frac{{r_0}}{r^2 (r+1) \log ({r_0}+1)}\right)^{m+n}+\beta  n\right)+1\right)}{r^2 (r+1) \log ({r_0}+1)} \nonumber\\
&-&
\left \lvert 2^{-n-3} ((r+1) \log (r+1)-r) \left(\frac{{r_0}}{r^2 (r+1) \log ({r_0}+1)}\right)^{-n-1}\right.\nonumber\\
&\times& \bigg[\alpha  (m-1) m \left(-2^{m+n}\right) \left(\frac{{r_0}}{r^2 (r+1) \log ({r_0}+1)}\right)^{m+n}
+\beta  n^2+\beta  n\bigg]
\nonumber\\
&-&\frac{1}{r^3 \log ({r_0}+1)}{r_0}\log (r+1)
\times\bigg[2^{-n-1} \left(\alpha  m 2^{m+n} \left(\frac{{r_0}}{r^2 (r+1) \log ({r_0}+1)}\right)^{m+n}+\beta  n\right)
\nonumber\\
&+&
\left(\frac{{r_0}}{r^2 (r+1) \log ({r_0}+1)}\right)^{-n-1}\bigg]
-\frac{1}{{r_0}^3}2^{-n-3}r^5(r+1)^3
\nonumber
\end{eqnarray}

\begin{eqnarray}
	&\times&\left(m \left(m^2-3 m+2\right)\alpha 2^{m+n}+\beta n^3+3 \beta  n^2+2 \beta  n
	\left(\frac{r_0}{r^2 (r+1) \log ({r_0}+1)}\right)^{m+n}\right)\nonumber\\
	&\times&\left.\left(\frac{r_0}{r^2 (r+1) \log ({r_0}+1)}\right)^{-n}\log ^2({r_0}+1)r \log ({r_0}+1)-{r_0} \log (r+1)\right \rvert
\end{eqnarray}

\begin{eqnarray}
\rho-\lvert p_t \rvert&=&
\frac{{r_0} \left(2^{-n-1} \left(\frac{{r_0}}{r^2 (r+1) \log ({r_0}+1)}\right)^{-n-1} \left(\alpha  m 2^{m+n} \left(\frac{{r_0}}{r^2 (r+1) \log ({r_0}+1)}\right)^{m+n}+\beta  n\right)+1\right)}{r^2 (r+1) \log ({r_0}+1)} \nonumber\\
&-&
\left \lvert-\frac{1}{2 r^3 (r+1) \log ({r_0}+1)}\bigg[{r_0}r-(r+1) \log (r+1)\bigg[(1+2^{-n-1} \bigg[(\alpha  m 2^{m+n}\right. \nonumber\\
&\times&\left(\frac{{r_0}}{r^2 (r+1) \log ({r_0}+1)}\right)^{m+n}+\beta  n\bigg]\left(\frac{{r_0}}{r^2 (r+1) \log ({r_0}+1)}\right)^{-n-1}\bigg]\nonumber\\
&+&2^{-n-1} r (r+1) \bigg[\alpha  (m-1) m \left(-2^{m+n}\right) \left(\frac{{r_0}}{r^2 (r+1) \log ({r_0}+1)}\right)^{m+n}+\beta  n^2 \nonumber\\
&+&\left.\beta  n\bigg]({r_0} \log (r+1)-r \log ({r_0}+1)) \left(\frac{{r_0}}{r^2 (r+1) \log ({r_0}+1)}\right)^{-n-2}\bigg]\right \rvert
\end{eqnarray}

\begin{eqnarray}
p_t-p_r&=&-\frac{1}{2 r^3 (r+1) \log ({r_0}+1)}\bigg[{r_0}r-(r+1) \log (r+1)\bigg[(1+2^{-n-1} \bigg[(\alpha  m 2^{m+n} \nonumber\\
&\times&\left(\frac{{r_0}}{r^2 (r+1) \log ({r_0}+1)}\right)^{m+n}+\beta  n\bigg]\left(\frac{{r_0}}{r^2 (r+1) \log ({r_0}+1)}\right)^{-n-1}\bigg]\nonumber\\
&+&2^{-n-1} r (r+1) \bigg[\alpha  (m-1) m \left(-2^{m+n}\right) \left(\frac{{r_0}}{r^2 (r+1) \log ({r_0}+1)}\right)^{m+n}+\beta  n^2 \nonumber\\
&+&\beta  n\bigg]({r_0} \log (r+1)-r \log ({r_0}+1)) \left(\frac{{r_0}}{r^2 (r+1) \log ({r_0}+1)}\right)^{-n-2}\bigg]
\nonumber\\
&+&
\frac{1}{2 r^3 (r+1) \log ({r_0}+1)}\bigg[{r_0}r-(r+1) \log (r+1)\bigg[(1+2^{-n-1} \bigg[(\alpha  m 2^{m+n} \nonumber\\
&\times&\left(\frac{{r_0}}{r^2 (r+1) \log ({r_0}+1)}\right)^{m+n}+\beta  n\bigg]\left(\frac{{r_0}}{r^2 (r+1) \log ({r_0}+1)}\right)^{-n-1}\bigg]\nonumber\\
&+&2^{-n-1} r (r+1) \bigg[\alpha  (m-1) m \left(-2^{m+n}\right) \left(\frac{{r_0}}{r^2 (r+1) \log ({r_0}+1)}\right)^{m+n}+\beta  n^2 \nonumber\\
&+&\beta  n\bigg]({r_0} \log (r+1)-r \log ({r_0}+1)) \left(\frac{{r_0}}{r^2 (r+1) \log ({r_0}+1)}\right)^{-n-2}\bigg]
\end{eqnarray}

\begin{figure}
	\centering
	\subfigure[WEC]{\includegraphics[scale=.3]{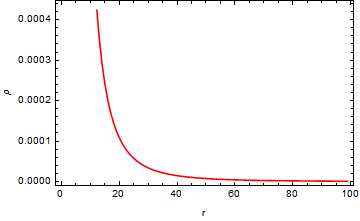}}\hspace{.05cm}
	\subfigure[NEC]{\includegraphics[scale=.3]{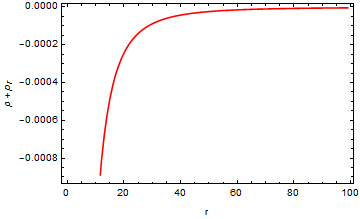}}\hspace{.05cm}
	\subfigure[NEC]{\includegraphics[scale=.3]{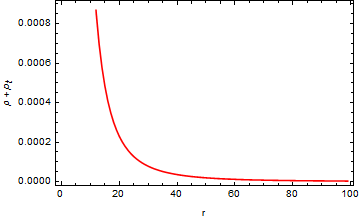}}\hspace{.05cm}
	\subfigure[SEC]{\includegraphics[scale=.3]{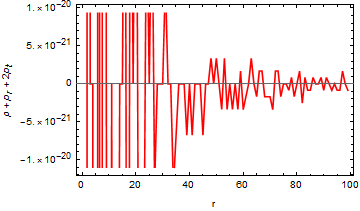}}\hspace{.05cm}
	\subfigure[DEC]{\includegraphics[scale=.3]{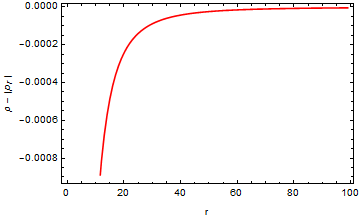}}\hspace{.05cm}
	\subfigure[DEC]{\includegraphics[scale=.3]{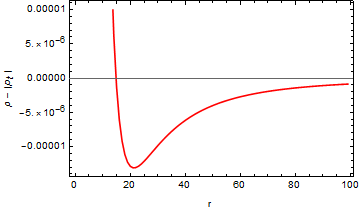}}\hspace{.05cm}
	\subfigure[$\omega$]{\includegraphics[scale=.3]{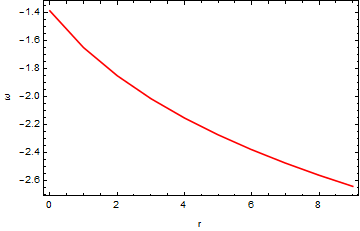}}\hspace{.05cm}
	\subfigure[$\triangle$]{\includegraphics[scale=.3]{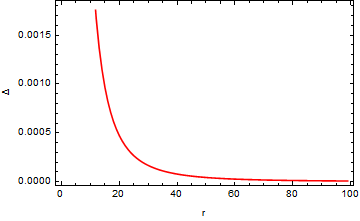}}\hspace{.05cm}
	\caption{Subcase 1(i): Plots for NEC, WEC, SEC, DEC, $\omega$ \& $\triangle$ with $\alpha=0$, $\beta=0$ }
\end{figure}

\begin{figure}
	\centering
	\subfigure[WEC]{\includegraphics[scale=.3]{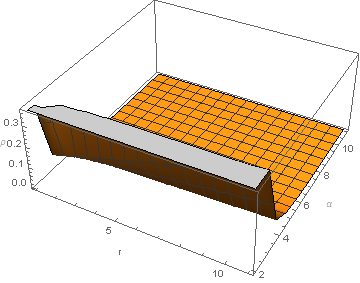}}\hspace{.05cm}
	\subfigure[NEC]{\includegraphics[scale=.3]{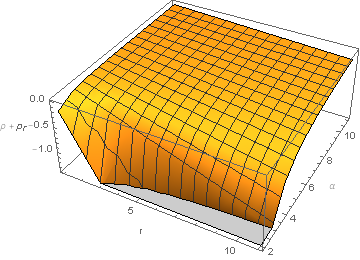}}\hspace{.05cm}
	\subfigure[NEC]{\includegraphics[scale=.3]{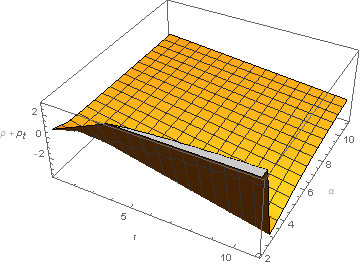}}\hspace{.05cm}
	\subfigure[SEC]{\includegraphics[scale=.3]{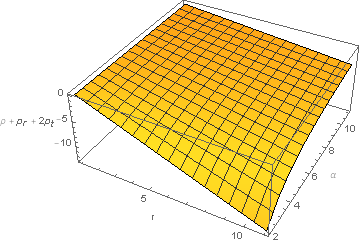}}\hspace{.05cm}
	\subfigure[DEC]{\includegraphics[scale=.3]{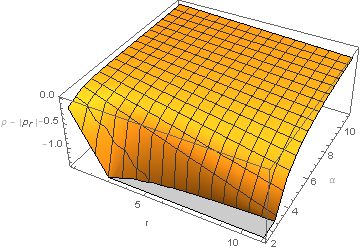}}\hspace{.05cm}
	\subfigure[DEC]{\includegraphics[scale=.3]{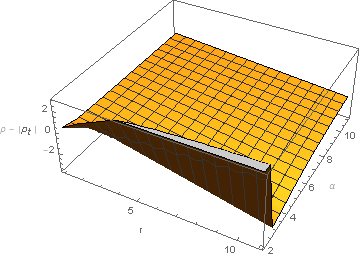}}\vspace{.05cm}
	\subfigure[$\omega$]{\includegraphics[scale=.3]{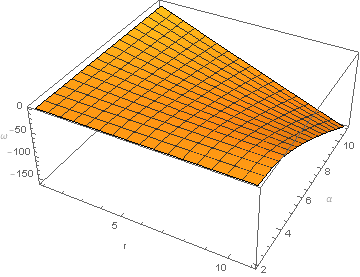}}\hspace{.05cm}
	\subfigure[$\triangle$]{\includegraphics[scale=.3]{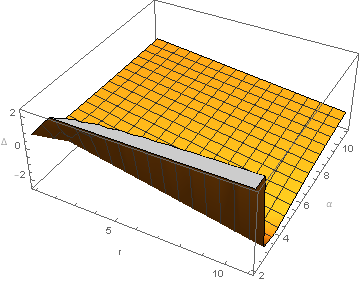}}\hspace{.05cm}
	\caption{Subcase 1(ii): Plots for NEC, WEC, SEC, DEC, $\omega$ \& $\triangle$ with  $\beta=0$}
\end{figure}

\begin{figure}
	\centering
	\subfigure[WEC]{\includegraphics[scale=.3]{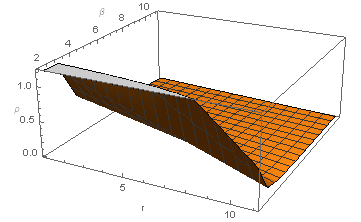}}\hspace{.05cm}
	\subfigure[NEC]{\includegraphics[scale=.3]{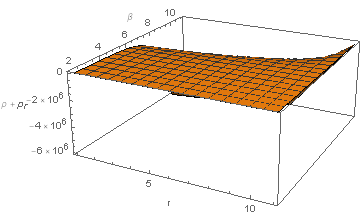}}\hspace{.05cm}
	\subfigure[NEC]{\includegraphics[scale=.3]{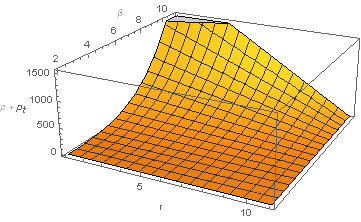}}\hspace{.05cm}
	\subfigure[SEC]{\includegraphics[scale=.3]{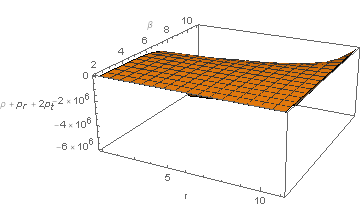}}\hspace{.05cm}
	\subfigure[DEC]{\includegraphics[scale=.3]{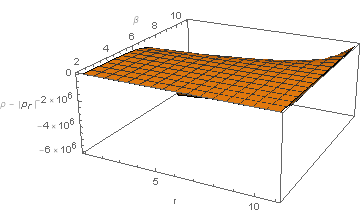}}\hspace{.05cm}
	\subfigure[DEC]{\includegraphics[scale=.3]{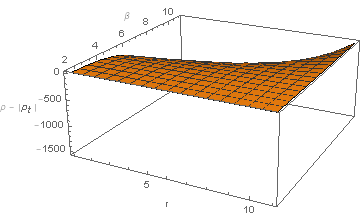}}\vspace{.05cm}
	\subfigure[$\omega$]{\includegraphics[scale=.3]{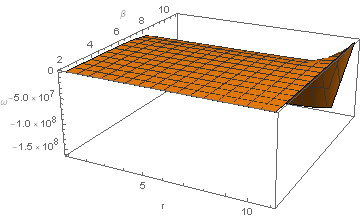}}\hspace{.05cm}
	\subfigure[$\triangle$]{\includegraphics[scale=.3]{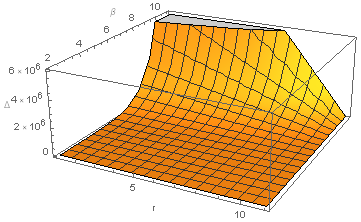}}\hspace{.05cm}
		\caption{Subcase 1(iii): Plots for NEC, WEC, SEC, DEC, $\omega$ \& $\triangle$ with $\alpha=0$}
\end{figure}

\begin{figure}
	\centering
	\subfigure[WEC]{\includegraphics[scale=.3]{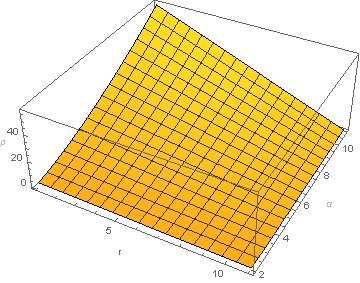}}\hspace{.05cm}
	\subfigure[NEC]{\includegraphics[scale=.3]{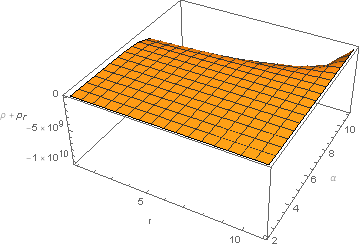}}\hspace{.05cm}
	\subfigure[NEC]{\includegraphics[scale=.3]{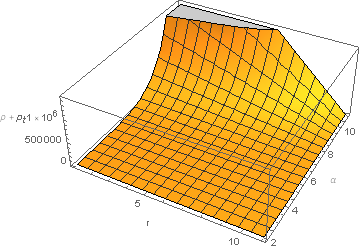}}\hspace{.05cm}
	\subfigure[SEC]{\includegraphics[scale=.3]{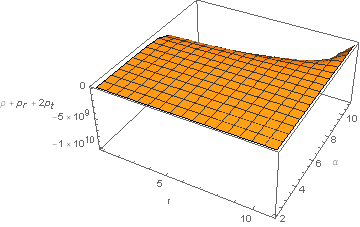}}\hspace{.05cm}
	\subfigure[DEC]{\includegraphics[scale=.3]{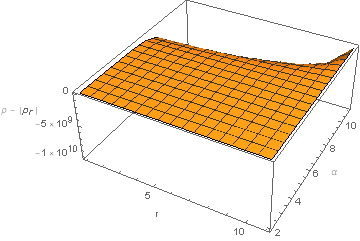}}\hspace{.05cm}
	\subfigure[DEC]{\includegraphics[scale=.3]{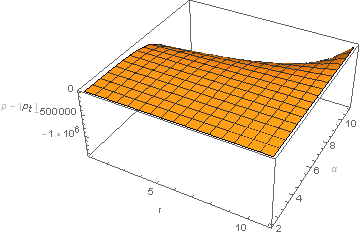}}\vspace{.05cm}
	\subfigure[$\omega$]{\includegraphics[scale=.3]{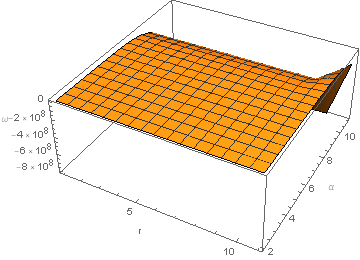}}\hspace{.05cm}
	\subfigure[$\triangle$]{\includegraphics[scale=.3]{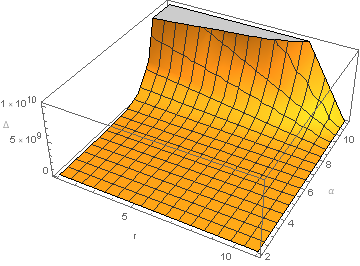}}\hspace{.05cm}
	\caption{Subcase 1(iv): Plots for NEC, WEC, SEC, DEC, $\omega$ \& $\triangle$ with $\beta=-0.5$}
\end{figure}

\begin{figure}
	\centering
	\subfigure[WEC]{\includegraphics[scale=.3]{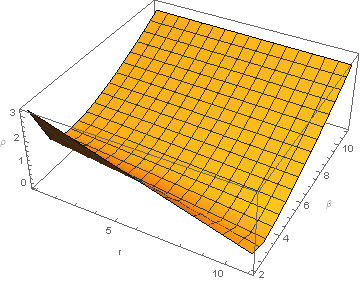}}\hspace{.05cm}
	\subfigure[NEC]{\includegraphics[scale=.3]{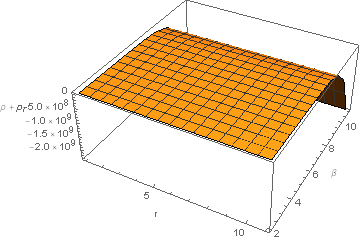}}\hspace{.05cm}
	\subfigure[NEC]{\includegraphics[scale=.3]{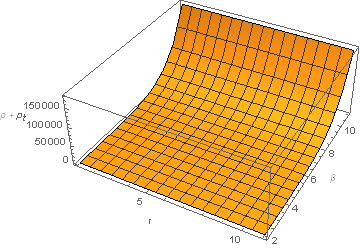}}\hspace{.05cm}
	\subfigure[SEC]{\includegraphics[scale=.3]{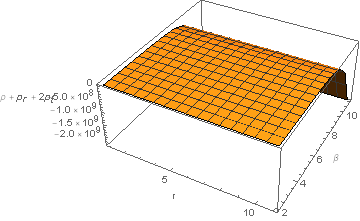}}\hspace{.05cm}
	\subfigure[DEC]{\includegraphics[scale=.3]{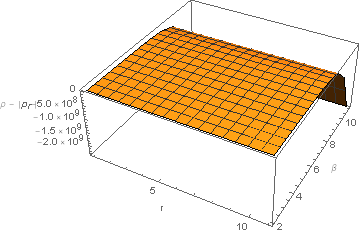}}\hspace{.05cm}
	\subfigure[DEC]{\includegraphics[scale=.3]{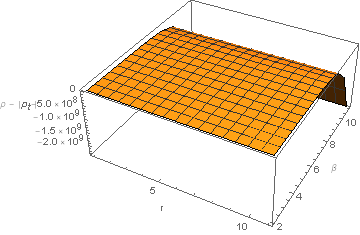}}\vspace{.05cm}
	\subfigure[$\omega$]{\includegraphics[scale=.3]{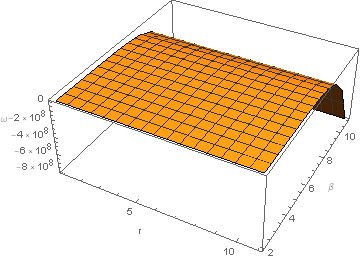}}\hspace{.05cm}
	\subfigure[$\triangle$]{\includegraphics[scale=.3]{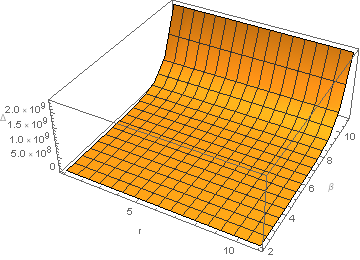}}\hspace{.05cm}
	\caption{Subcase 1(v): Plots for NEC, WEC, SEC, DEC, $\omega$ \& $\triangle$ with $\alpha=-0.5$}
\end{figure}

\subsection{Case II}
In this case, the wormhole solutions are obtained by using the shape function $b(r)=r_0(\frac{r}{r_0})^\gamma$, $0<\gamma<1$. The energy density, radial pressure and tangential pressure  are obtained from the field equations \eqref{6} - \eqref{8} as follows:
\begin{eqnarray}\label{20}
\rho &=&\frac{1}{2} \left(\alpha  2^m m \left(-\frac{(\gamma -1) \left(\frac{{r_0}}{r}\right)^{\gamma }}{r^2}\right)^m+\beta  2^{-n} n \left(-\frac{(\gamma -1) \left(\frac{{r_0}}{r}\right)^{\gamma }}{r^2}\right)^{-n}-\frac{2 (\gamma -1) \left(\frac{{r_0}}{r}\right)^{\gamma }}{r^2}\right)
\end{eqnarray}

\begin{eqnarray}\label{21}
p_r&=&\frac{1}{(\gamma -1)^3 r^2}\bigg[2^{-n-3} \left(\frac{{r_0}}{r}\right)^{-3 \gamma } \left(-\frac{(\gamma -1) \left(\frac{{r_0}}{r}\right)^{\gamma }}{r^2}\right)^{-n} \left(2^n \left(-\frac{(\gamma -1) \left(\frac{{r_0}}{r}\right)^{\gamma }}{r^2}\right)^n \left(-\alpha  2^m m^3 r^8\right.\right.\nonumber \\
&\times&
 \left.\left.\left(\left(\frac{{r_0}}{r}\right)^{\gamma }-1\right) \left(-\frac{(\gamma -1) \left(\frac{{r_0}}{r}\right)^{\gamma }}{r^2}\right)^m+3 \alpha  2^m m^2 r^8 \left(\left(\frac{{r_0}}{r}\right)^{\gamma }-1\right) \left(-\frac{(\gamma -1) \left(\frac{{r_0}}{r}\right)^{\gamma }}{r^2}\right)^m\right.\right.\nonumber \\
  &+&\left.\left.\gamma ^2 \left(\frac{{r_0}}{r}\right)^{2 \gamma } \left(\alpha  \left(-2^m\right) m^2 r^5 \left(-\frac{(\gamma -1) \left(\frac{{r_0}}{r}\right)^{\gamma }}{r^2}\right)^m+\alpha  2^m m r^2 \left(r^3+4 \left(\frac{{r_0}}{r}\right)^{\gamma }\right)\right.\right. \right.\nonumber \\
  &\times&
  \left.\left.\left.\left(-\frac{(\gamma -1) \left(\frac{{r_0}}{r}\right)^{\gamma }}{r^2}\right)^m+24 \left(\frac{{r_0}}{r}\right)^{2 \gamma }\right)-\gamma  \left(\frac{{r_0}}{r}\right)^{2 \gamma } \left(\alpha  \left(-2^m\right) m^2 r^5 \left(-\frac{(\gamma -1) \left(\frac{{r_0}}{r}\right)^{\gamma }}{r^2}\right)^m\right.\right. \right.\nonumber \\
  &+&  \left.\left.\left.\alpha  2^m m r^2 \left(r^3+8 \left(\frac{{r_0}}{r}\right)^{\gamma }\right) \left(-\frac{(\gamma -1) \left(\frac{{r_0}}{r}\right)^{\gamma }}{r^2}\right)^m+24 \left(\frac{{r_0}}{r}\right)^{2 \gamma }\right)+\alpha  2^{m+1} m r^2 \left(2 \left(\frac{{r_0}}{r}\right)^{3 \gamma }\right.\right. \right.\nonumber\\
  &-&\left.\left.\left.r^6 \left(\left(\frac{{r_0}}{r}\right)^{\gamma }-1\right)\right) \left(-\frac{(\gamma -1) \left(\frac{{r_0}}{r}\right)^{\gamma }}{r^2}\right)^m-8 \gamma ^3 \left(\frac{{r_0}}{r}\right)^{4 \gamma }+8 \left(\frac{{r_0}}{r}\right)^{4 \gamma }\right)-\beta  n^3 r^8 \left(\left(\frac{{r_0}}{r}\right)^{\gamma }\right. \right.\nonumber
   \end{eqnarray}
  \begin{eqnarray}
  &-&\left.\left.1\right)-\beta  n^2 r^5 \left(3 r^3 \left(\left(\frac{{r_0}}{r}\right)^{\gamma }-1\right)-(\gamma -1) \gamma  \left(\frac{{r_0}}{r}\right)^{2 \gamma }\right)+\beta  n r^2 \left(-2 r^6 \left(\left(\frac{{r_0}}{r}\right)^{\gamma }-1\right)\right. \right.\nonumber \\
  &+&  \left.\left.(\gamma -1) \gamma  r^3 \left(\frac{{r_0}}{r}\right)^{2 \gamma }+4 (\gamma -1)^2 \left(\frac{{r_0}}{r}\right)^{3 \gamma }\right)\right)\bigg]
\end{eqnarray}

\begin{eqnarray}\label{22}
p_t&=&\frac{1}{4 (\gamma -1)^2 r^3}\bigg[2^{-n} r^6 \left(1-\left(\frac{{r_0}}{r}\right)^{\gamma }\right) \left(\frac{{r_0}}{r}\right)^{-2 \gamma } \left(-\frac{(\gamma -1) \left(\frac{{r_0}}{r}\right)^{\gamma }}{r^2}\right)^{-n} \left(\alpha  (m-1) m \left(-2^{m+n}\right)\right.\nonumber \\
&\times&
\left. \left(-\frac{(\gamma -1) \left(\frac{{r_0}}{r}\right)^{\gamma }}{r^2}\right)^{m+n}+\beta  n^2+\beta  n\right)+(1-\gamma ) \gamma  r \left(\frac{{r_0}}{r}\right)^{\gamma } \left(-2 \gamma +\alpha  (\gamma -1) \left(-2^m\right) m\right. \nonumber \\
&\times&
\left.\left(-\frac{(\gamma -1) \left(\frac{{r_0}}{r}\right)^{\gamma }}{r^2}\right)^{m-1}-\beta  (\gamma -1) 2^{-n} n \left(-\frac{(\gamma -1) \left(\frac{{r_0}}{r}\right)^{\gamma }}{r^2}\right)^{-n-1}+2\right)\bigg]
\end{eqnarray}

Adding equations \eqref{20} and \eqref{21}, we have

\begin{eqnarray}
\rho + p_r&=&\frac{1}{2} \left(\alpha  2^m m \left(-\frac{(\gamma -1) \left(\frac{{r_0}}{r}\right)^{\gamma }}{r^2}\right)^m+\frac{\beta n}{2^n} \left(-\frac{(\gamma -1) \left(\frac{{r_0}}{r}\right)^{\gamma }}{r^2}\right)^{-n}-\frac{2 (\gamma -1) \left(\frac{{r_0}}{r}\right)^{\gamma }}{r^2}\right)
\end{eqnarray}

\begin{eqnarray}
&+&\frac{1}{(\gamma -1)^3 r^2}\bigg[2^{-n-3} \left(\frac{{r_0}}{r}\right)^{-3 \gamma } \left(-\frac{(\gamma -1) \left(\frac{{r_0}}{r}\right)^{\gamma }}{r^2}\right)^{-n} \left(2^n \left(-\frac{(\gamma -1) \left(\frac{{r_0}}{r}\right)^{\gamma }}{r^2}\right)^n \left(-\alpha  2^m m^3 r^8\right.\right.\nonumber \\
&\times&
\left.\left.\left(\left(\frac{{r_0}}{r}\right)^{\gamma }-1\right) \left(-\frac{(\gamma -1) \left(\frac{{r_0}}{r}\right)^{\gamma }}{r^2}\right)^m+3 \alpha  2^m m^2 r^8 \left(\left(\frac{{r_0}}{r}\right)^{\gamma }-1\right) \left(-\frac{(\gamma -1) \left(\frac{{r_0}}{r}\right)^{\gamma }}{r^2}\right)^m\right.\right.\nonumber \\
&+&\left.\left.\gamma ^2 \left(\frac{{r_0}}{r}\right)^{2 \gamma } \left(\alpha  \left(-2^m\right) m^2 r^5 \left(-\frac{(\gamma -1) \left(\frac{{r_0}}{r}\right)^{\gamma }}{r^2}\right)^m+\alpha  2^m m r^2 \left(r^3+4 \left(\frac{{r_0}}{r}\right)^{\gamma }\right)\right.\right. \right.\nonumber \\
&\times&
\left.\left.\left.\left(-\frac{(\gamma -1) \left(\frac{{r_0}}{r}\right)^{\gamma }}{r^2}\right)^m+24 \left(\frac{{r_0}}{r}\right)^{2 \gamma }\right)-\gamma  \left(\frac{{r_0}}{r}\right)^{2 \gamma } \left(\alpha  \left(-2^m\right) m^2 r^5 \left(-\frac{(\gamma -1) \left(\frac{{r_0}}{r}\right)^{\gamma }}{r^2}\right)^m\right.\right. \right.\nonumber \\
&+&  \left.\left.\left.\alpha  2^m m r^2 \left(r^3+8 \left(\frac{{r_0}}{r}\right)^{\gamma }\right) \left(-\frac{(\gamma -1) \left(\frac{{r_0}}{r}\right)^{\gamma }}{r^2}\right)^m+24 \left(\frac{{r_0}}{r}\right)^{2 \gamma }\right)+\alpha  2^{m+1} m r^2 \left(2 \left(\frac{{r_0}}{r}\right)^{3 \gamma }\right.\right. \right.\nonumber \\
&-&\left.\left.\left.r^6 \left(\left(\frac{{r_0}}{r}\right)^{\gamma }-1\right)\right) \left(-\frac{(\gamma -1) \left(\frac{{r_0}}{r}\right)^{\gamma }}{r^2}\right)^m-8 \gamma ^3 \left(\frac{{r_0}}{r}\right)^{4 \gamma }+8 \left(\frac{{r_0}}{r}\right)^{4 \gamma }\right)-\beta  n^3 r^8 \left(\left(\frac{{r_0}}{r}\right)^{\gamma }\right. \right.\nonumber \\
&-&\left.\left.1\right)-\beta  n^2 r^5 \left(3 r^3 \left(\left(\frac{{r_0}}{r}\right)^{\gamma }-1\right)-(\gamma -1) \gamma  \left(\frac{{r_0}}{r}\right)^{2 \gamma }\right)+\beta  n r^2 \left(-2 r^6 \left(\left(\frac{{r_0}}{r}\right)^{\gamma }-1\right)\right. \right.\nonumber \\
&+&  \left.\left.(\gamma -1) \gamma  r^3 \left(\frac{{r_0}}{r}\right)^{2 \gamma }+4 (\gamma -1)^2 \left(\frac{{r_0}}{r}\right)^{3 \gamma }\right)\right)\bigg]
\end{eqnarray}
Adding equations \eqref{20} and \eqref{22}, we have

\begin{eqnarray}
\rho+p_t &=&\frac{1}{2} \left(\alpha  2^m m \left(-\frac{(\gamma -1) \left(\frac{{r_0}}{r}\right)^{\gamma }}{r^2}\right)^m+\beta  2^{-n} n \left(-\frac{(\gamma -1) \left(\frac{{r_0}}{r}\right)^{\gamma }}{r^2}\right)^{-n}-\frac{2 (\gamma -1) \left(\frac{{r_0}}{r}\right)^{\gamma }}{r^2}\right)\nonumber \\
&+&\frac{1}{4 (\gamma -1)^2 r^3}\bigg[2^{-n} r^6 \left(1-\left(\frac{{r_0}}{r}\right)^{\gamma }\right) \left(\frac{{r_0}}{r}\right)^{-2 \gamma } \left(-\frac{(\gamma -1) \left(\frac{{r_0}}{r}\right)^{\gamma }}{r^2}\right)^{-n} \left(\alpha  (m-1) m \left(-2^{m+n}\right)\right.\nonumber \\
&\times&
\left. \left(-\frac{(\gamma -1) \left(\frac{{r_0}}{r}\right)^{\gamma }}{r^2}\right)^{m+n}+\beta  n^2+\beta  n\right)+(1-\gamma ) \gamma  r \left(\frac{{r_0}}{r}\right)^{\gamma } \left(-2 \gamma +\alpha  (\gamma -1) \left(-2^m\right) m\right. \nonumber \\
&\times&
\left.\left(-\frac{(\gamma -1) \left(\frac{{r_0}}{r}\right)^{\gamma }}{r^2}\right)^{m-1}-\beta  (\gamma -1) 2^{-n} n \left(-\frac{(\gamma -1) \left(\frac{{r_0}}{r}\right)^{\gamma }}{r^2}\right)^{-n-1}+2\right)\bigg]
\end{eqnarray}
Adding equations \eqref{20}, \eqref{21} and $2\times \eqref{22}$, we have
\begin{eqnarray}
\rho + p_r+2p_t&=&\frac{1}{2} \left(\frac{\alpha m}{2^{-m}} \left(\frac{(1-\gamma) \left(\frac{{r_0}}{r}\right)^{\gamma }}{r^2}\right)^m+\frac{\beta n}{2^n} \left(\frac{(1-\gamma) \left(\frac{{r_0}}{r}\right)^{\gamma }}{r^2}\right)^{-n}-\frac{2 (\gamma -1) \left(\frac{{r_0}}{r}\right)^{\gamma }}{r^2}\right)
\nonumber
\end{eqnarray}

\begin{eqnarray}
&+&\frac{1}{(\gamma -1)^3 r^2}\bigg[2^{-n-3} \left(\frac{{r_0}}{r}\right)^{-3 \gamma } \left(-\frac{(\gamma -1) \left(\frac{{r_0}}{r}\right)^{\gamma }}{r^2}\right)^{-n} \left(2^n \left(-\frac{(\gamma -1) \left(\frac{{r_0}}{r}\right)^{\gamma }}{r^2}\right)^n \left(-\alpha  2^m m^3 r^8\right.\right.\nonumber \\
&\times&
\left.\left.\left(\left(\frac{{r_0}}{r}\right)^{\gamma }-1\right) \left(-\frac{(\gamma -1) \left(\frac{{r_0}}{r}\right)^{\gamma }}{r^2}\right)^m+3 \alpha  2^m m^2 r^8 \left(\left(\frac{{r_0}}{r}\right)^{\gamma }-1\right) \left(-\frac{(\gamma -1) \left(\frac{{r_0}}{r}\right)^{\gamma }}{r^2}\right)^m\right.\right.\nonumber \\
&+&\left.\left.\gamma ^2 \left(\frac{{r_0}}{r}\right)^{2 \gamma } \left(\alpha  \left(-2^m\right) m^2 r^5 \left(-\frac{(\gamma -1) \left(\frac{{r_0}}{r}\right)^{\gamma }}{r^2}\right)^m+\alpha  2^m m r^2 \left(r^3+4 \left(\frac{{r_0}}{r}\right)^{\gamma }\right)\right.\right. \right.\nonumber \\
&\times&
\left.\left.\left.\left(-\frac{(\gamma -1) \left(\frac{{r_0}}{r}\right)^{\gamma }}{r^2}\right)^m+24 \left(\frac{{r_0}}{r}\right)^{2 \gamma }\right)-\gamma  \left(\frac{{r_0}}{r}\right)^{2 \gamma } \left(\alpha  \left(-2^m\right) m^2 r^5 \left(-\frac{(\gamma -1) \left(\frac{{r_0}}{r}\right)^{\gamma }}{r^2}\right)^m\right.\right. \right.\nonumber \\
&+&  \left.\left.\left.\alpha  2^m m r^2 \left(r^3+8 \left(\frac{{r_0}}{r}\right)^{\gamma }\right) \left(-\frac{(\gamma -1) \left(\frac{{r_0}}{r}\right)^{\gamma }}{r^2}\right)^m+24 \left(\frac{{r_0}}{r}\right)^{2 \gamma }\right)+\alpha  2^{m+1} m r^2 \left(2 \left(\frac{{r_0}}{r}\right)^{3 \gamma }\right.\right. \right.\nonumber \\
&-&\left.\left.\left.r^6 \left(\left(\frac{{r_0}}{r}\right)^{\gamma }-1\right)\right) \left(-\frac{(\gamma -1) \left(\frac{{r_0}}{r}\right)^{\gamma }}{r^2}\right)^m-8 \gamma ^3 \left(\frac{{r_0}}{r}\right)^{4 \gamma }+8 \left(\frac{{r_0}}{r}\right)^{4 \gamma }\right)-\beta  n^3 r^8 \left(\left(\frac{{r_0}}{r}\right)^{\gamma }\right. \right.\nonumber \\
&-&\left.\left.1\right)-\beta  n^2 r^5 \left(3 r^3 \left(\left(\frac{{r_0}}{r}\right)^{\gamma }-1\right)-(\gamma -1) \gamma  \left(\frac{{r_0}}{r}\right)^{2 \gamma }\right)+\beta  n r^2 \left(-2 r^6 \left(\left(\frac{{r_0}}{r}\right)^{\gamma }-1\right)\right. \right.\nonumber \\
&+&  \left.\left.(\gamma -1) \gamma  r^3 \left(\frac{{r_0}}{r}\right)^{2 \gamma }+4 (\gamma -1)^2 \left(\frac{{r_0}}{r}\right)^{3 \gamma }\right)\right)\bigg]\nonumber
\end{eqnarray}
  \begin{eqnarray}
&+&2\bigg[\frac{1}{4 (\gamma -1)^2 r^3}\bigg[2^{-n} r^6 \left(1-\left(\frac{{r_0}}{r}\right)^{\gamma }\right) \left(\frac{{r_0}}{r}\right)^{-2 \gamma } \left(-\frac{(\gamma -1) \left(\frac{{r_0}}{r}\right)^{\gamma }}{r^2}\right)^{-n} \left(\alpha  (m-1) m \left(-2^{m+n}\right)\right.\nonumber\\
&\times&
\left. \left(-\frac{(\gamma -1) \left(\frac{{r_0}}{r}\right)^{\gamma }}{r^2}\right)^{m+n}+\beta  n^2+\beta  n\right)+(1-\gamma ) \gamma  r \left(\frac{{r_0}}{r}\right)^{\gamma } \left(-2 \gamma +\alpha  (\gamma -1) \left(-2^m\right) m\right. \nonumber \\
&\times&
\left.\left(-\frac{(\gamma -1) \left(\frac{{r_0}}{r}\right)^{\gamma }}{r^2}\right)^{m-1}-\beta  (\gamma -1) 2^{-n} n \left(-\frac{(\gamma -1) \left(\frac{{r_0}}{r}\right)^{\gamma }}{r^2}\right)^{-n-1}+2\right)\bigg]\bigg]
\end{eqnarray}

\begin{eqnarray}
\rho -\lvert p_r\rvert&=&\frac{1}{2} \left(\alpha  2^m m \left(-\frac{(\gamma -1) \left(\frac{{r_0}}{r}\right)^{\gamma }}{r^2}\right)^m+\beta  2^{-n} n \left(-\frac{(\gamma -1) \left(\frac{{r_0}}{r}\right)^{\gamma }}{r^2}\right)^{-n}-\frac{2 (\gamma -1) \left(\frac{{r_0}}{r}\right)^{\gamma }}{r^2}\right)\nonumber \\
&-&\bigg\lvert\frac{1}{(\gamma -1)^3 r^2}\bigg[2^{-n-3} \left(\frac{{r_0}}{r}\right)^{-3 \gamma } \left(-\frac{(\gamma -1) \left(\frac{{r_0}}{r}\right)^{\gamma }}{r^2}\right)^{-n} \left(2^n \left(-\frac{(\gamma -1) \left(\frac{{r_0}}{r}\right)^{\gamma }}{r^2}\right)^n \left(-\alpha  2^m m^3 r^8\right.\right.\nonumber \\
&\times&
\left.\left.\left(\left(\frac{{r_0}}{r}\right)^{\gamma }-1\right) \left(-\frac{(\gamma -1) \left(\frac{{r_0}}{r}\right)^{\gamma }}{r^2}\right)^m+3 \alpha  2^m m^2 r^8 \left(\left(\frac{{r_0}}{r}\right)^{\gamma }-1\right) \left(-\frac{(\gamma -1) \left(\frac{{r_0}}{r}\right)^{\gamma }}{r^2}\right)^m\right.\right.\nonumber \\
&+&\left.\left.\gamma ^2 \left(\frac{{r_0}}{r}\right)^{2 \gamma } \left(\alpha  \left(-2^m\right) m^2 r^5 \left(-\frac{(\gamma -1) \left(\frac{{r_0}}{r}\right)^{\gamma }}{r^2}\right)^m+\alpha  2^m m r^2 \left(r^3+4 \left(\frac{{r_0}}{r}\right)^{\gamma }\right)\right.\right. \right.\nonumber
\end{eqnarray}

\begin{eqnarray}
&\times&
\left.\left.\left.\left(-\frac{(\gamma -1) \left(\frac{{r_0}}{r}\right)^{\gamma }}{r^2}\right)^m+24 \left(\frac{{r_0}}{r}\right)^{2 \gamma }\right)-\gamma  \left(\frac{{r_0}}{r}\right)^{2 \gamma } \left(\alpha  \left(-2^m\right) m^2 r^5 \left(-\frac{(\gamma -1) \left(\frac{{r_0}}{r}\right)^{\gamma }}{r^2}\right)^m\right.\right. \right.\nonumber \\
&+&  \left.\left.\left.\alpha  2^m m r^2 \left(r^3+8 \left(\frac{{r_0}}{r}\right)^{\gamma }\right) \left(-\frac{(\gamma -1) \left(\frac{{r_0}}{r}\right)^{\gamma }}{r^2}\right)^m+24 \left(\frac{{r_0}}{r}\right)^{2 \gamma }\right)+\alpha  2^{m+1} m r^2 \left(2 \left(\frac{{r_0}}{r}\right)^{3 \gamma }\right.\right. \right.\nonumber \\
&-&\left.\left.\left.r^6 \left(\left(\frac{{r_0}}{r}\right)^{\gamma }-1\right)\right) \left(-\frac{(\gamma -1) \left(\frac{{r_0}}{r}\right)^{\gamma }}{r^2}\right)^m-8 \gamma ^3 \left(\frac{{r_0}}{r}\right)^{4 \gamma }+8 \left(\frac{{r_0}}{r}\right)^{4 \gamma }\right)-\beta  n^3 r^8 \left(\left(\frac{{r_0}}{r}\right)^{\gamma }\right. \right.\nonumber \\
&-&\left.\left.1\right)-\beta  n^2 r^5 \left(3 r^3 \left(\left(\frac{{r_0}}{r}\right)^{\gamma }-1\right)-(\gamma -1) \gamma  \left(\frac{{r_0}}{r}\right)^{2 \gamma }\right)+\beta  n r^2 \left(-2 r^6 \left(\left(\frac{{r_0}}{r}\right)^{\gamma }-1\right)\right. \right.\nonumber \\
&+&  \left.\left.(\gamma -1) \gamma  r^3 \left(\frac{{r_0}}{r}\right)^{2 \gamma }+4 (\gamma -1)^2 \left(\frac{{r_0}}{r}\right)^{3 \gamma }\right)\right)\bigg]\bigg\rvert
\end{eqnarray}

\begin{eqnarray}
\rho-\lvert p_t\rvert &=&\frac{1}{2} \left(\alpha  2^m m \left(-\frac{(\gamma -1) \left(\frac{{r_0}}{r}\right)^{\gamma }}{r^2}\right)^m+\beta  2^{-n} n \left(-\frac{(\gamma -1) \left(\frac{{r_0}}{r}\right)^{\gamma }}{r^2}\right)^{-n}-\frac{2 (\gamma -1) \left(\frac{{r_0}}{r}\right)^{\gamma }}{r^2}\right)\nonumber \\
&-&\bigg\lvert\frac{1}{4 (\gamma -1)^2 r^3}\bigg[2^{-n} r^6 \left(1-\left(\frac{{r_0}}{r}\right)^{\gamma }\right) \left(\frac{{r_0}}{r}\right)^{-2 \gamma } \left(-\frac{(\gamma -1) \left(\frac{{r_0}}{r}\right)^{\gamma }}{r^2}\right)^{-n} \left(\alpha  (m-1) m \left(-2^{m+n}\right)\right.\nonumber \\
&\times&
\left. \left(-\frac{(\gamma -1) \left(\frac{{r_0}}{r}\right)^{\gamma }}{r^2}\right)^{m+n}+\beta  n^2+\beta  n\right)+(1-\gamma ) \gamma  r \left(\frac{{r_0}}{r}\right)^{\gamma } \left(-2 \gamma +\alpha  (\gamma -1) \left(-2^m\right) m\right. \nonumber \\
&\times&
\left.\left(-\frac{(\gamma -1) \left(\frac{{r_0}}{r}\right)^{\gamma }}{r^2}\right)^{m-1}-\beta  (\gamma -1) 2^{-n} n \left(-\frac{(\gamma -1) \left(\frac{{r_0}}{r}\right)^{\gamma }}{r^2}\right)^{-n-1}+2\right)\bigg]\bigg\rvert
\end{eqnarray}

\begin{eqnarray}
p_t-p_r&=&\frac{1}{4 (\gamma -1)^2 r^3}\bigg[2^{-n} r^6 \left(1-\left(\frac{{r_0}}{r}\right)^{\gamma }\right) \left(\frac{{r_0}}{r}\right)^{-2 \gamma } \left(-\frac{(\gamma -1) \left(\frac{{r_0}}{r}\right)^{\gamma }}{r^2}\right)^{-n} \left(\alpha  (m-1) m \left(-2^{m+n}\right)\right.\nonumber \\
&\times&
\left. \left(-\frac{(\gamma -1) \left(\frac{{r_0}}{r}\right)^{\gamma }}{r^2}\right)^{m+n}+\beta  n^2+\beta  n\right)+(1-\gamma ) \gamma  r \left(\frac{{r_0}}{r}\right)^{\gamma } \left(-2 \gamma +\alpha  (\gamma -1) \left(-2^m\right) m\right. \nonumber \\
&\times&
\left.\left(-\frac{(\gamma -1) \left(\frac{{r_0}}{r}\right)^{\gamma }}{r^2}\right)^{m-1}-\beta  (\gamma -1) 2^{-n} n \left(-\frac{(\gamma -1) \left(\frac{{r_0}}{r}\right)^{\gamma }}{r^2}\right)^{-n-1}+2\right)\bigg]\nonumber \\
&-&\bigg[\frac{1}{(\gamma -1)^3 r^2}\bigg[2^{-n-3} \left(\frac{{r_0}}{r}\right)^{-3 \gamma } \left(-\frac{(\gamma -1) \left(\frac{{r_0}}{r}\right)^{\gamma }}{r^2}\right)^{-n} \left(2^n \left(-\frac{(\gamma -1) \left(\frac{{r_0}}{r}\right)^{\gamma }}{r^2}\right)^n \left(-\alpha  2^m m^3 r^8\right.\right.\nonumber
 \\
&\times&
\left.\left.\left(\left(\frac{{r_0}}{r}\right)^{\gamma }-1\right) \left(-\frac{(\gamma -1) \left(\frac{{r_0}}{r}\right)^{\gamma }}{r^2}\right)^m+3 \alpha  2^m m^2 r^8 \left(\left(\frac{{r_0}}{r}\right)^{\gamma }-1\right) \left(-\frac{(\gamma -1) \left(\frac{{r_0}}{r}\right)^{\gamma }}{r^2}\right)^m\right.\right.\nonumber
\end{eqnarray}

\begin{eqnarray}
&+&\left.\left.\gamma ^2 \left(\frac{{r_0}}{r}\right)^{2 \gamma } \left(\alpha  \left(-2^m\right) m^2 r^5 \left(-\frac{(\gamma -1) \left(\frac{{r_0}}{r}\right)^{\gamma }}{r^2}\right)^m+\alpha  2^m m r^2 \left(r^3+4 \left(\frac{{r_0}}{r}\right)^{\gamma }\right)\right.\right. \right.\nonumber \\
&\times&
\left.\left.\left.\left(-\frac{(\gamma -1) \left(\frac{{r_0}}{r}\right)^{\gamma }}{r^2}\right)^m+24 \left(\frac{{r_0}}{r}\right)^{2 \gamma }\right)-\gamma  \left(\frac{{r_0}}{r}\right)^{2 \gamma } \left(\alpha  \left(-2^m\right) m^2 r^5 \left(-\frac{(\gamma -1) \left(\frac{{r_0}}{r}\right)^{\gamma }}{r^2}\right)^m\right.\right. \right.\nonumber \\
&+&  \left.\left.\left.\alpha  2^m m r^2 \left(r^3+8 \left(\frac{{r_0}}{r}\right)^{\gamma }\right) \left(-\frac{(\gamma -1) \left(\frac{{r_0}}{r}\right)^{\gamma }}{r^2}\right)^m+24 \left(\frac{{r_0}}{r}\right)^{2 \gamma }\right)+\alpha  2^{m+1} m r^2 \left(2 \left(\frac{{r_0}}{r}\right)^{3 \gamma }\right.\right. \right.\nonumber \\
&-&\left.\left.\left.r^6 \left(\left(\frac{{r_0}}{r}\right)^{\gamma }-1\right)\right) \left(-\frac{(\gamma -1) \left(\frac{{r_0}}{r}\right)^{\gamma }}{r^2}\right)^m-8 \gamma ^3 \left(\frac{{r_0}}{r}\right)^{4 \gamma }+8 \left(\frac{{r_0}}{r}\right)^{4 \gamma }\right)-\beta  n^3 r^8 \left(\left(\frac{{r_0}}{r}\right)^{\gamma }\right. \right.\nonumber \\
&-&\left.\left.1\right)-\beta  n^2 r^5 \left(3 r^3 \left(\left(\frac{{r_0}}{r}\right)^{\gamma }-1\right)-(\gamma -1) \gamma  \left(\frac{{r_0}}{r}\right)^{2 \gamma }\right)+\beta  n r^2 \left(-2 r^6 \left(\left(\frac{{r_0}}{r}\right)^{\gamma }-1\right)\right. \right.\nonumber \\
&+&  \left.\left.(\gamma -1) \gamma  r^3 \left(\frac{{r_0}}{r}\right)^{2 \gamma }+4 (\gamma -1)^2 \left(\frac{{r_0}}{r}\right)^{3 \gamma }\right)\right)\bigg]\bigg]
\end{eqnarray}

\begin{figure}
	\centering
	\subfigure[WEC]{\includegraphics[scale=.3]{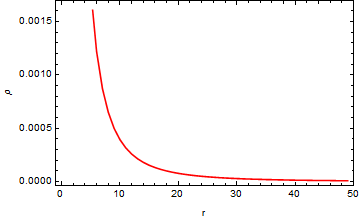}}\hspace{.05cm}
	\subfigure[NEC]{\includegraphics[scale=.3]{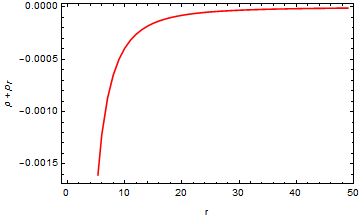}}\hspace{.05cm}
	\subfigure[NEC]{\includegraphics[scale=.3]{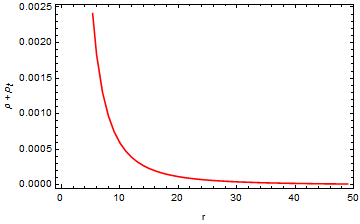}}\hspace{.05cm}
	\subfigure[SEC]{\includegraphics[scale=.3]{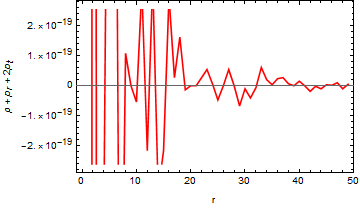}}\hspace{.05cm}
	\subfigure[DEC]{\includegraphics[scale=.3]{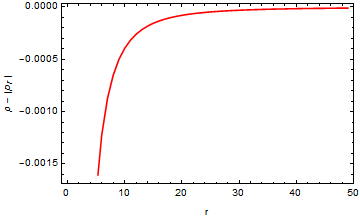}}\hspace{.05cm}
	\subfigure[DEC]{\includegraphics[scale=.3]{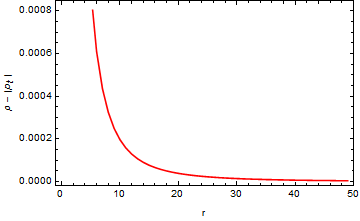}}\hspace{.05cm}
		\subfigure[$\omega$]{\includegraphics[scale=.3]{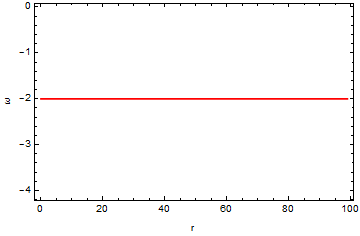}}\hspace{.05cm}
	\subfigure[$\triangle$]{\includegraphics[scale=.3]{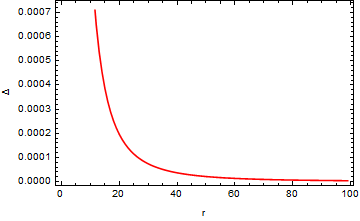}}\hspace{.05cm}
	\caption{Subcase 2(i): Plots for NEC, WEC, SEC \& DEC, $\omega$ \& $\triangle$ with $\alpha=0$, $\beta=0$ }
\end{figure}

\begin{figure}
	\centering
	\subfigure[WEC]{\includegraphics[scale=.3]{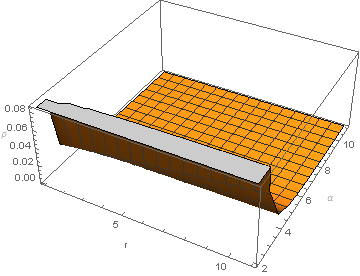}}\hspace{.05cm}
	\subfigure[NEC]{\includegraphics[scale=.3]{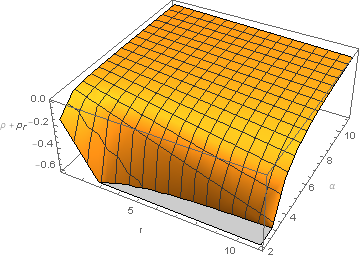}}\hspace{.05cm}
	\subfigure[NEC]{\includegraphics[scale=.3]{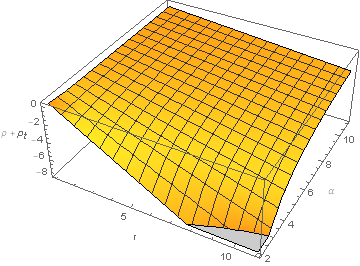}}\hspace{.05cm}
	\subfigure[SEC]{\includegraphics[scale=.3]{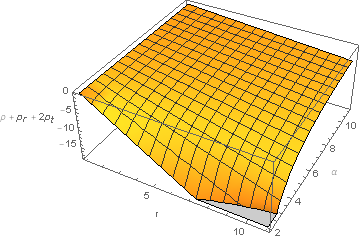}}\hspace{.05cm}
	\subfigure[DEC]{\includegraphics[scale=.3]{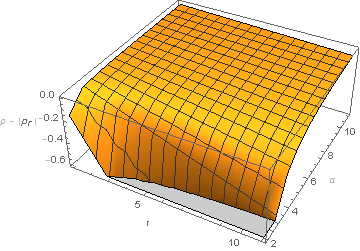}}\hspace{.05cm}
	\subfigure[DEC]{\includegraphics[scale=.3]{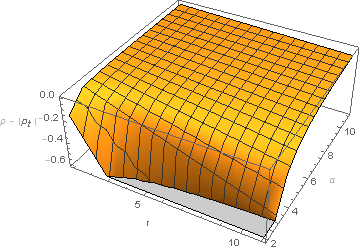}}\hspace{.05cm}
	\subfigure[$\omega$]{\includegraphics[scale=.3]{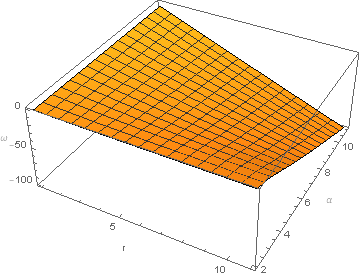}}\hspace{.05cm}
	\subfigure[$\triangle$]{\includegraphics[scale=.3]{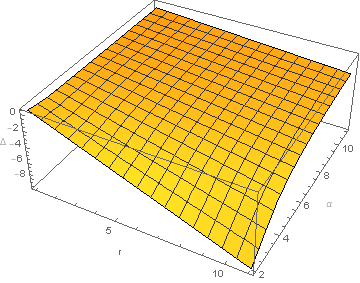}}\hspace{.05cm}
	\caption{Subcase 2(ii): Plots for NEC, WEC, SEC, DEC, $\omega$ \& $\triangle$ with $\beta=0$ }
\end{figure}

\begin{figure}
	\centering
	\subfigure[WEC]{\includegraphics[scale=.3]{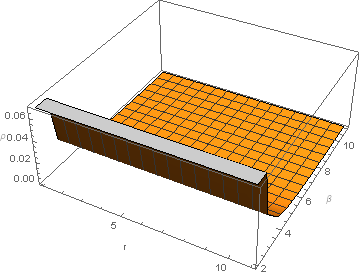}}\hspace{.05cm}
	\subfigure[NEC]{\includegraphics[scale=.3]{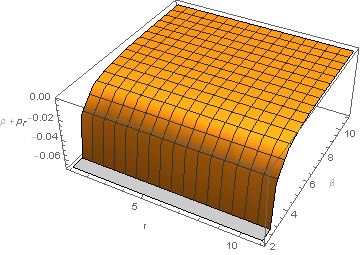}}\hspace{.05cm}
	\subfigure[NEC]{\includegraphics[scale=.3]{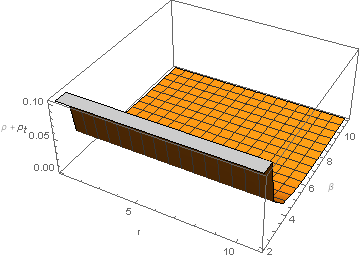}}\hspace{.05cm}
	\subfigure[SEC]{\includegraphics[scale=.3]{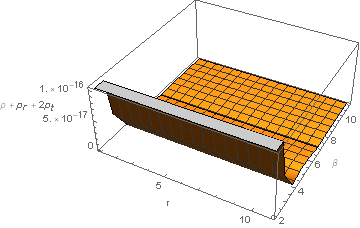}}\hspace{.05cm}
	\subfigure[DEC]{\includegraphics[scale=.3]{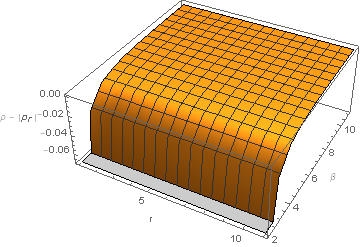}}\hspace{.05cm}
	\subfigure[DEC]{\includegraphics[scale=.3]{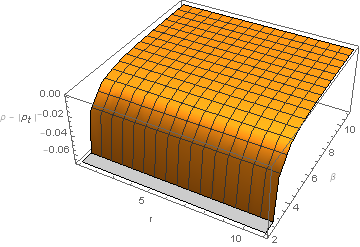}}\hspace{.05cm}
	\subfigure[$\omega$]{\includegraphics[scale=.3]{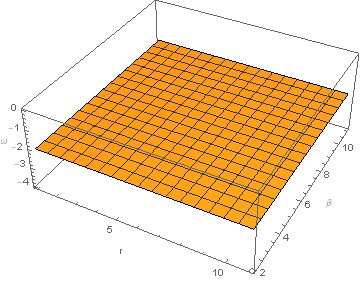}}\hspace{.05cm}
	\subfigure[$\triangle$]{\includegraphics[scale=.3]{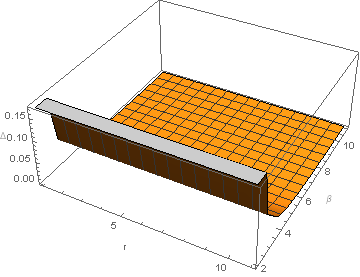}}\hspace{.05cm}
	\caption{Subcase 2(iii): Plots for NEC, WEC, SEC, DEC, $\omega$ \& $\triangle$ with $\alpha=0$ }
\end{figure}

\begin{figure}
	\centering
	\subfigure[WEC]{\includegraphics[scale=.3]{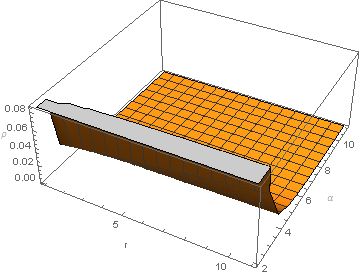}}\hspace{.05cm}
	\subfigure[NEC]{\includegraphics[scale=.3]{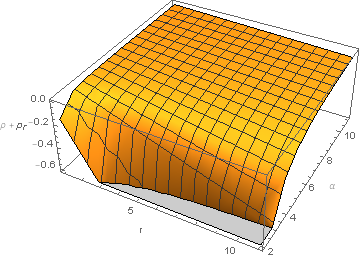}}\hspace{.05cm}
	\subfigure[NEC]{\includegraphics[scale=.3]{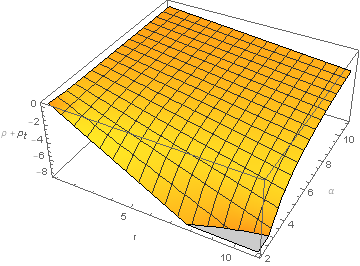}}\hspace{.05cm}
	\subfigure[SEC]{\includegraphics[scale=.3]{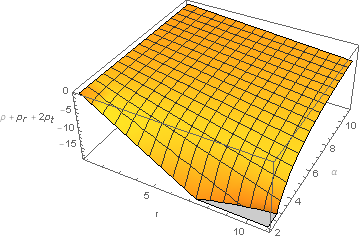}}\hspace{.05cm}
	\subfigure[DEC]{\includegraphics[scale=.3]{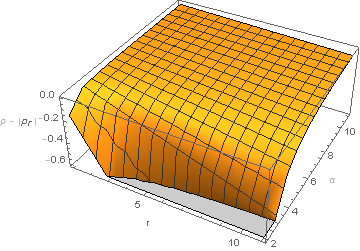}}\hspace{.05cm}
	\subfigure[DEC]{\includegraphics[scale=.3]{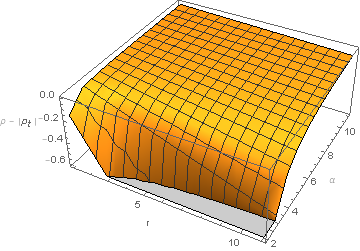}}\hspace{.05cm}
	\subfigure[$\omega$]{\includegraphics[scale=.3]{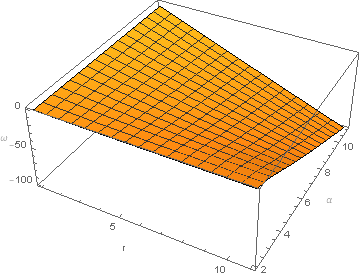}}\hspace{.05cm}
	\subfigure[$\triangle$]{\includegraphics[scale=.3]{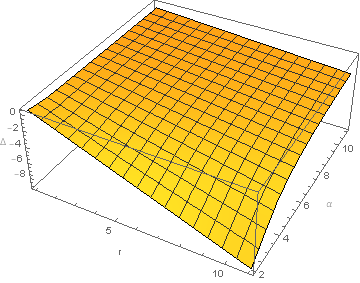}}\hspace{.05cm}
	\caption{Subcase 2(iv): Plots for NEC, WEC, SEC, DEC, $\omega$ \& $\triangle$ with $\beta=-0.5$ }
\end{figure}

\begin{figure}
	\centering
	\subfigure[WEC]{\includegraphics[scale=.3]{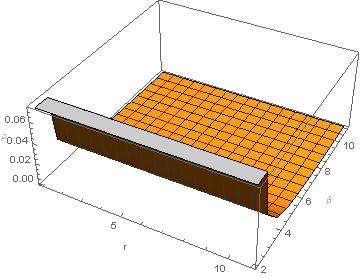}}\hspace{.05cm}
	\subfigure[NEC]{\includegraphics[scale=.3]{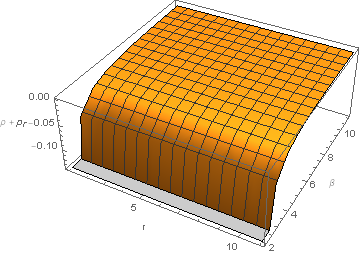}}\hspace{.05cm}
	\subfigure[NEC]{\includegraphics[scale=.3]{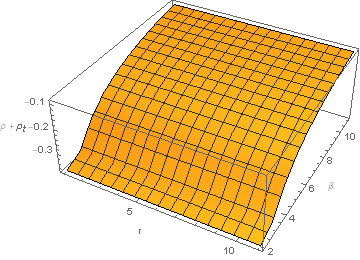}}\hspace{.05cm}
	\subfigure[SEC]{\includegraphics[scale=.3]{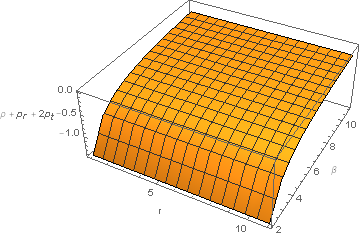}}\hspace{.05cm}
	\subfigure[DEC]{\includegraphics[scale=.3]{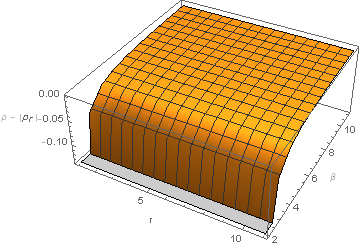}}\hspace{.05cm}
	\subfigure[DEC]{\includegraphics[scale=.3]{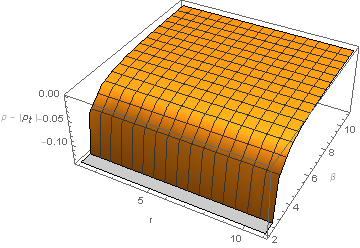}}\hspace{.05cm}
	\subfigure[$\omega$]{\includegraphics[scale=.3]{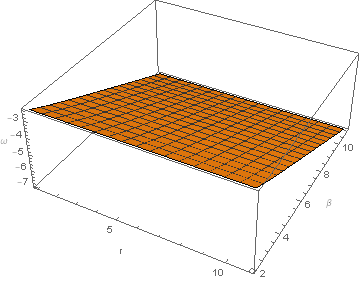}}\hspace{.05cm}
	\subfigure[$\triangle$]{\includegraphics[scale=.3]{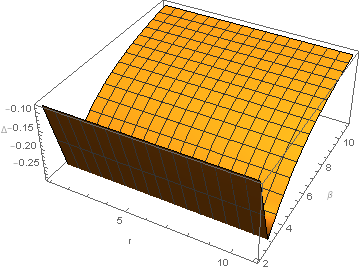}}\hspace{.05cm}
		\caption{Subcase 2(v): Plots for NEC, WEC, SEC, DEC, $\omega$ \& $\triangle$ with $\alpha=0.5$}
\end{figure}

\section{Results and Perspectives}
The important energy conditions are the Null Energy Condition (NEC), Weak Energy Condition (WEC), Strong Energy Condition (SEC) and Dominated Energy Condition (DEC). These conditions are expressed as follows:
\begin{itemize}
  \item [(I)] $\rho + p_r\geq 0$, $\rho + p_t\geq 0$
  \item [(II)] $\rho \geq 0$, $\rho + p_r> 0$, $\rho + p_t> 0$
  \item [(III)] $\rho + p_r\geq 0$, $\rho + p_t\geq 0$,  $\rho + p_r +2p_t\geq 0$
  \item [(IV)] $\rho\ge 0$, $\rho - \lvert p_r\rvert \geq 0$, $\rho - \lvert p_t\rvert \geq 0$
\end{itemize}
The wormholes are non-vacuum solutions of Einstein's field equations and according to Einstein's field theory, they are filled with a matter which is different from the normal matter and is known as exotic matter. This matter does not validate the energy conditions.

In this paper, the wormhole solutions are explored with two  different shape functions in $f(R)$ theory of gravity. The function $f(R)$ is taken as $f(R)=R+\alpha R^m-\beta R^{-n}$, where $m$ and $n$ are arbitrary constants. The two shape functions are chosen as  $b(r)=\dfrac{{r_0} \log (r+1)}{\log ({r_0}+1)}$ and $b(r)=r_0(\frac{r}{r_0})^\gamma$. Both of the functions satisfy all the conditions discussed in Section 2. In this work, the energy conditions are investigated and, anisotropy parameter \& equation of state parameter are analyzed based on two different shape functions in two cases. Since the function $f(R)$ contains two parameters
$\alpha$ and $\beta$, so  according to the values of these parameters, both of the cases have  been dealt with the following five subcases: (i) $\alpha=0$, $\beta=0$,  (ii)  	$\alpha$-variable, $\beta=0$, (iii) 	$\alpha=0$,		$\beta$-variable, (iv) 	$\alpha$-variable, $\beta$-non-zero constant, (v)	$\alpha$-non-zero constant,	$\beta$-variable.

For each shape function, Subcase (i) reduces to the general relativity, Subcase (ii), reduces to the Starobinsky model \cite{star} with $m=2$ which was  introduced for the explanation of early time inflation, Subcase (iii) leads to the form $f(R)=R-\beta R^{-n}$ used by Cao et al. \cite{cao} for the investigation of late time acceleration in the context of FRW universe,  Subcase (iv) converts to   $f(R)=R+\alpha R^{-\frac{1}{2}}-\beta R^{\frac{1}{2}}$ with $m =n=-\dfrac{1}{2}$, $\alpha$ as variable and $\beta$ as a non-zero constant, studied in \cite{cao}, finally in Subcase (v) it takes the form   $f(R)=R+\alpha R^{-\frac{1}{2}}-\beta R^{\frac{1}{2}}$ for case 1 and    $f(R)=R+\alpha R^{\frac{1}{2}}-\beta R^{\frac{1}{2}}$ for case 2 with $\alpha$ as constant and $\beta$ as variable. The energy density $\rho$, various combinations of energy density and pressure i.e. $\rho + p_r$, $\rho + p_t$,  $\rho + p_r +2p_t$, $\rho - \lvert p_r\rvert$ and $\rho - \lvert p_t\rvert$, equation of state parameter and anisotropy parameter are plotted for all five subcases of Case 1 in Figures 1-5 and for Case 2 in Figures 6-10  with respect to the radial coordinate $r$.

For Case 1, in Subcase 1(i), the function adopts the form $f(R)=R$, which is  free from parameters $\alpha$ and $\beta$. Therefore, two dimensional graphs are obtained in Fig. (1). In rest subcases, either $\alpha$ is variable or $\beta$ is variable. That is why, three dimensional graphs are obtained in all other four subcases. In Subcase 1(i), $\rho$ and $\rho + p_t$ are obtained to be positive and decreasing to zero with the variation of $r$ (Figures 1(a) \& 1(c)). However, $\rho + p_r$ and  $\rho - \lvert p_r\rvert$ are come out be negative, increasing and tending towards zero with the increment in $r$ (Figures 1(b) and 1(e)). One DEC term  $\rho - \lvert p_t\rvert$  is decreasing for small range of $r$ and then it is increasing, negative and tending to zero (Fig. 1(f)). One SEC term $\rho + p_r +2p_t$  is fluctuating between positive and negative values and going to zero with the increment of $r$ (Fig. 1(d)). The anisotropy parameter is obtained to be positive (Fig. 1(h)). Thus, all the energy conditions are violated and geometry is repulsive in Subcase 1(i).
In Subcase 1(ii),  $f(R)=R+\alpha R^2$.  Varying the parameter $\alpha$ along with  $r$, only the energy density is obtained to be positive and tending to zero (Fig. 2(a)). All other energy condition terms are found to negative with the variation of $\alpha$ and $r$ (Figures 2(b)-2(f)). The anisotropy parameter possesses negative values with respect to radial coordinate (Fig. (2(h))).  This subcase shows the violation of all energy conditions and the geometry to be of attractive in nature.
In Subcase 1(iii), $f(R)=R-\beta R^{-n}$. Taking $n=-0.5$ and varying $\beta$ and $r$, $\rho$ and $\rho +p_t$ are found to be positively decreasing functions of $r$ (Figures 3(a), 3(c)). Other energy condition terms are found to be negatively decreasing functions of $r$ (Figures 3(b), 3(d)-3(f)).  The anisotropy parameter is negative value with respect to radial coordinate (Fig. 3(h)). This shows attractive nature of geometry and violation of all energy conditions.
In Subcases 1(iv) and 1(v),  the form of $f(R)$ is $f(R)=R+\alpha R^{-\frac{1}{2}}-\beta R^{\frac{1}{2}}$.
In Subcase 1(iv), taking $\beta=-0.5$ and  $\alpha$ as variable, $\rho$ and $\rho + p_t$ are found to be positive and decreasing functions with respect to $r$ (Figures 4(a), 4(c)). All other combinations of pressure and energy density are found to be negative (Figures 4(b), 4(d)-4(f)). In Subcase (v), $\alpha$ is taken to be equal to -0.5 and $\beta$ is varied. Then, again both  $\rho$ and $\rho + p_t$ are found to be positive and decreasing functions with respect to $r$ (Figures 5(a), 5(c)) and other energy condition terms are come out to be negative (Figures 5(b), 5(d), 5(e), 5(f)). The anisotropy parameter has positive values with the increment in radial coordinate in both subcases 1(iv) and 1(v) (Figures 4(h), 5(h)). This represents that the geometry is repulsive in nature and all energy conditions are violated in both Subcases 1(iv) and 1(v).

For Case 2, in Subcase 2(i), the graphs are drawn for $f(R)=R$. In Figures 6(a), 6(c) \& 6(f), $\rho$, $\rho + p_t$ and $\rho - \lvert p_t\rvert$ are positive, decreasing and tending to zero as $r$ tends to infinity. In Figures 6(b) and 6(e), $\rho + p_r$ and  $\rho - \lvert p_r\rvert$ are found to be negative, increasing and tending towards zero with the increment in $r$. Like Subcase 1(i), SEC term $\rho + p_r +2p_t$  fluctuates between positive and negative values and then tends to zero as $r$ increases (Fig. 6(d)). The anisotropy parameter is come out to be positive with radial coordinate (Fig. (6(h))). Thus, the geometry has repulsive nature and all the energy conditions are violated in Subcase 2(i).
In Subcase 2(ii),  for $f(R)=R+\alpha R^2$, the parameter $\alpha$ is varied with  $r$. In this subcase, only the energy density is obtained to be positively decreasing function and tending to zero (Fig. 7(a)). However, all other energy condition terms are found to be negative with the change in $\alpha$ and $r$ (Figures 7(b)-7(f)). The anisotropy parameter has negative values (Fig. (7(h))). Hence, the geometry is attractive and  all the energy conditions are dissatisfied.
In Subcase 2(iii), the function $f(R)$ is read as $f(R)=R-\beta R^{-n}$. Taking $n=-0.5$ and varying $\beta$ and $r$, $\rho$, $\rho +p_t$  \& $\rho +p_r+2p_t$ are found to be positively decreasing functions of $r$ (Figures 8(a), 8(c), 8(d)). Other energy condition terms are found to have negative values with the increment in  $r$ (Figures 8(b), 8(e), 8(f)). The anisotropy parameter is obtained to be positive (Fig. 8(h)). This again shows the violation of all energy conditions and the geometry to be repulsive in nature.
In Subcases 2(iv),  the function $f(R)$ has the form $f(R)=R+\alpha R^{-\frac{1}{2}}-\beta R^{\frac{1}{2}}$ with $\beta=-0.5$ and  $\alpha$ as a variable. Here, only $\rho$ possesses positive values and decreases with respect to $r$ (Figures 9(a)). However, all other combinations take negative values (Figures 9(b)-9(f)). The anisotropy parameter is obtained to negative with the radial coordinate (Fig. 9(h)). Thus, all energy conditions are dissatisfied and geometry is attractive in this subcase.
In Subcase 2(v), $f(R)=R+\alpha R^{\frac{1}{2}}-\beta R^{\frac{1}{2}}$ with $\alpha=0.5$ and  $\beta$ as a variable. Here also, only  $\rho$ is positive and decreasing functions of $r$ (Figure 10(a)) and other energy condition terms are negative (Figures 10(b)-10(f)). The anisotropy parameter is found  to have negative values with the change in radial coordinate (Fig. 10(h)). This represents an attractive nature of geometry and  violation of all energy conditions. In each subcase of both cases, the equation of state parameter is observed to have values less than -1 with the increment in the radial coordinate (Figures (1(g), 2(g), 3(g), 4(g), 5(g), 6(g), 7(g), 8(g), 9(g), 10(g))).
Since the energy density is obtained to be positive in every subcase (Figures (1(a), 2(a), 3(a), 4(a), 5(a), 6(a), 7(a), 8(a), 9(a), 10(a))), therefore the value of the mass function, defined in Eq. \ref{mass}, is also positive.
 Thus, no subcase satisfies any energy condition and hence strongly shows the presence of exotic matter which confirms the existence of wormholes in the universe.

In this work, wormhole models are presented in $f(R)$ gravity. The structure of wormholes is characterized by the choice of the shape function and this shape function needs to satisfy various conditions. Here, two shape functions satisfying all the desired properties are taken to investigate wormholes with the background of $f(R)$ model in which the function $f(R)$ is taken to be dependent on two parameters. Depending upon the values of these parameters, its five forms are dealt with for each type of shape function. The energy conditions are calculated and plotted. All these are found to be violated (Table 1). Since the existence of exotic or abnormal matter demands this violation, therefore the  framed wormholes strongly confirm their existence in the universe. The value of equation of parameter is obtained to be less than -1 which specify the wormholes to be filled with phantom fluid. In both cases, the value of anisotropy parameter, that shows the nature of geometry, is found to have same nature for GR and Starobinsky models for both types of shape functions. In GR case,  it shows repulsive nature, while in Starobinsky case, it represents attractive nature. In rest cases, the behavior of anisotropy parameter is different for different shape functions. Thus, every shape function has a significant role in determining the geometry, the type of filled fluid and the type of present matter in the context of wormholes.
\begin{table}[!h]
	\centering
	\begin{tabular}{|c|c|c|c|c|c|c|}
		\hline
		\multicolumn{7}{|c|}{$f(R)=R+\alpha R^m-\beta R^{-n}$ with (i)  $b(r)=\dfrac{{r_0} \log (r+1)}{\log ({r_0}+1)}$ \&  (ii) $b(r)=r_0(\frac{r}{r_0})^\gamma$}\\
		\hline
		S.No.& Energy&$\alpha=0$ & 	$\alpha$-variable&	$\alpha=0$ &		$\alpha$-variable & 	$\alpha$-non-zero constant\\
		&Conditions&	$\beta=0$ & 	$\beta=0$&	$\beta$-variable &		$\beta$-non-zero constant & 	$\beta$-variable\\
		
		\hline
		1 & NEC & Violated & Violated & Violated & Violated & Violated \\
		
		2 & WEC & Violated & Violated & Violated & Violated & Violated\\
		
		3 & SEC & Violated & Violated & Violated &  Violated & Violated\\
		
		4 & DEC & Violated & Violated & Violated & Violated & Violated\\
		\hline
	\end{tabular}
\end{table} \\ 

\textbf{Acknowledgment:} GCS thanks to IUCAA, Pune (India) for hospitality during an academic visit where a part of
this work is executed. GCS is also thankful to Council of Scientific and Industrial Research (CSIR), Govt. of
India, for providing support (Ref. No. 25(0260)/17/EMR-II) for carrying out the research work. We are also very much thankful to the reviewer and editor
for their constructive comments to improve the work significantly.


\begin{thebibliography}{ii}
\bibitem{morris1} M. S. Morris and K. S. Thorne, Am. J. Phys. \textbf{56} (1988) 395.
\bibitem{haw}S. W. Hawking, Phys. Rev. D \textbf{46} (1992) 603.
\bibitem{fried} J. L. Friedmann, K. Schleich and D. N. Witt, Phys. Rev. Lett. \textbf{71} (1993) 1486.
\bibitem{fro} V. Frolov and I. D. Novikov, Phys. Rev. D \textbf{48} (1993) 1607.
\bibitem{visshoch} M. Visser and D. Hochberg, in: Proceedings of the Haifa Workshop on the Internal
Structure of Black Holes and Spacetime Singularities, (Haifa, Israel, 1997).
\bibitem{flamm}L. Flamm, Phys. Z. \textbf{17} (1916) 448.
\bibitem{eins-ros}A. Einstein and N. Rosen, Ann. Phys. \textbf{2} (1935) 242.
\bibitem{wheel} J. A. Wheeler, Geometrodynamics, (Academic Press, New York, 1962).


\bibitem{morris2} M. S. Morris, K. S. Thorne and U. Yurtsever, Phys. Rev. Lett. \textbf{61} (1988) 1446.
\bibitem{visser} M. Visser, Lorentzian wormholes: from Einstein to Hawking, AIP Press, New York (1995).		
\bibitem{star}A. A. Starobinsky, Phys. Lett. B \textbf{91} (1980) 99.	
\bibitem{sotiriou} T. P. Sotiriou, Class. Quant. Grav. \textbf{23} (2006) 5117.
\bibitem{Nojiri}S. Nojiri and S. D. Odintsov, Phys. Rept. \textbf{505} (2011) 59.
\bibitem{huang}Q. G. Huang, JCAP \textbf{02} (2014) 035.


\bibitem{noj}S. Nojiri and S. D. Odintsov, Int. J. Geom. Methods Mod.
Phys., \textbf{4}, (2007) 115.

\bibitem{cog2008} G. Cognola, E. Elizalde, S. Nojiri, S. D. Odintsov, L. Sebastiani and S. Zerbini, Phys. Rev. D \textbf{77} (2008) 046009.

\bibitem{felice} A. D. Felice and  S. Tsujikawa, Living Rev. Rel. \textbf{13} (2010) 3.

\bibitem{Bamba4}  	
K. Bamba, Chao-Qiang Geng and Chung-Chi Lee, JCAP \textbf{1011} (2010) 001.


\bibitem{Bamba3}  	
K. Bamba, Chao-Qiang Geng and Chung-Chi Lee, Int. J. Mod. Phys. D \textbf{20} (2011) 1339.

\bibitem{Bamba1}  	
K. Bamba, S. Nojiri and S. D. Odintsov, Phys. Rev. D \textbf{85} (2012) 044012.


\bibitem{sebas} 
L. Sebastiani, G. Cognola, R. Myrzakulov, S. D. Odintsov and S. Zerbini, Phys. Rev. D \textbf{89} (2014) 023518.

\bibitem{thakur}S. Thakur and A. A. Sen, Phys. Rev. D, \textbf{88} (2013) 044043.






\bibitem{Bamba}  	
K. Bamba, S. Nojiri, S. D. Odintsov and D. Saez-Gomez, Phys. Rev. D \textbf{90} (2014) 124061.

\bibitem{Peng}  	
P. Wang, H. Yang and S. Ying, Phys. Rev. D \textbf{96} (2017) 046007.


\bibitem{Motohashi}  	
H. Motohashi and A. A. Starobinsky, Eur. Phys. J. C \textbf{77} (2017) 538.

\bibitem{Astashenok}  	
A. V. Astashenok, S. D. Odintsov and A. de la Cruz-Dombriz, Class. Quant. Grav. \textbf{34} (2017) 205008.

\bibitem{noj2017} S. Nojiri, S.D. Odintsov and V.K. Oikonomou, Phys. Rept. \textbf{692} (2017) 1.


\bibitem{baha1}S. Bahamonde, S. D. Odintsov, V. K. Oikonomou and P. V.Tretyakov, Phys. Lett. B \textbf{766} (2017) 225.



\bibitem{Yousaf}  	
Z. Yousaf, K. Bamba and M. Zaeem-ul-Haq Bhatti, Phys. Rev. D \textbf{95} (2017) 024024.

\bibitem{Faraoni}  	
V. Faraoni and S. D. Belknap-Keet, Phys. Rev. D \textbf{96} (2017) 044040.

\bibitem{Abbas}  	
G. Abbas, M. S. Khan, Z. Ahmad and M. Zubair, Eur. Phys. J. C \textbf{77} (2017) 443.


\bibitem{Sussman}  	
R. A. Sussman and L. G. Jaime, Class. Quant. Grav. \textbf{34} (2017) 245004.


\bibitem{Mongwane}  	
B. Mongwane, Phys. Rev. D \textbf{96} (2017) 024028.


\bibitem{Mansour}  	
H. Mansour, B. S. Lakhal and A. Yanallah, JCAP \textbf{1806} (2018) 006.

\bibitem{Muller}  	
D. Muller, A. Ricciardone, A. A. Starobinsky and A. Toporensky,  Eur. Phys. J. C \textbf{78} (2018) 311.

\bibitem{Wang}  	
J. Wang, R. Gui and W. Qiu, Phys. Dark Univ. \textbf{19} (2018) 60.


\bibitem{Papagiannopoulos}  	
G. Papagiannopoulos, S. Basilakos, John D. Barrow and A. Paliathanasis, Phys. Rev. D \textbf{97} (2018) 024026.

\bibitem{Oikonomou}  	
V. K. Oikonomou, Phys. Rev. D \textbf{97} (2018) 064001.

\bibitem{Chakraborty}  	
S. Chakraborty, Phys. Rev. D \textbf{98} (2018) 024009.

\bibitem{Nashed} 	
G. G. L. Nashed, Int. J. Mod. Phys. D \textbf{27} (2018) 1850074.

\bibitem{Capozziello} 
S. Capozziello, S. Nojiri and S. D. Odintsov, Phys. Lett. B \textbf{781} (2018) 99.

\bibitem{Gu}  	
B. M. Gu, Y. X. Liu and Y. Zhong, Phys. Rev. D \textbf{98} (2018) 024027.

\bibitem{Abbas1} 	
G. Abbas and H. Nazar, Eur. Phys. J. C \textbf{78} (2018) 510.

\bibitem{Mishra}  	
A. K. Mishra, M. Rahman and S. Sarkar, Class. Quant. Grav. \textbf{35} (2018) 145011.

\bibitem{Odintsov}  	
S. D. Odintsov and V. K. Oikonomou, Phys. Rev. D \textbf{98} (2018) 024013.


\bibitem{Hochberg}
D. Hochberg, A. Popov and S. V. Sushkov, Phys. Rev. Lett. \textbf{78} (1997) 2050.

\bibitem{Nojiri1} S. Nojiri, O. Obregon, S. D. Odintsov and K.E. Osetrin, Phys. Lett. B \textbf{449} (1999) 173.

\bibitem{Nojiri2} S. Nojiri, O. Obregon, S. D. Odintsov and K.E. Osetrin, Phys. Lett. B \textbf{458} (1999) 19.




\bibitem{lemos} J. P. S. Lemos, F. S. N. Lobo  and S. Q. D. Oliveira, Phys. Rev. D \textbf{68} (2003) 064004.	

\bibitem{furey} N. Furey and A. De Benedictis, Class. Quant. Grav. \textbf{22} (2005) 313.


\bibitem{dotti} G. Dotti, J. Oliva and R. Troncoso, Phys. Rev. D \textbf{75} (2007) 024002.
\bibitem{lobo1} F. S. N. Lobo, Phys. Rev. D \textbf{73} (2006) 064028.
\bibitem{chakra} S. Chakraborty and T. Bandyopadhyay, Int. J. Mod. Phys. D \textbf{18} (2009) 463.
\bibitem{jamil} M. Jamil, M. U. Farooq and M. A. Rashid, Eur. Phys. J. C \textbf{59} (2009) 907.


\bibitem{lobo} F. S. N. Lobo and M. A. Oliveira, Phys. Rev. D \textbf{80} (2009) 104012.
\bibitem{cata} M. Cataldo, P. Meza and P. Minning, Phys. Rev. D \textbf{83} (2011) 044050.


\bibitem{saiedi} H. Saiedi and B. N. Esfahani, Mod. Phys. Lett. A \textbf{26} (2011) 1211.

\bibitem{lopez} M. Bouhmadi-L\'{o}pez, F. S. N. Lobo and P. Mart\'{i}n-Moruno, JCAP \textbf{1411} (2014) 007.
\bibitem{najafi} S. Najafi, T. Rostami and S. Jalalzadeh, Annals of Physics, \textbf{354} (2015) 288.
\bibitem{baha} S. Bahamonde, M. Jamil, P. Pavlovic and M. Sossich, Phys. Rev. D \textbf{94}, (2016) 044041.

\bibitem{Rahaman}  	
F. Rahaman, N. Paul, A. Banerjee, S. S. De, S. Ray and A. A. Usmani, Eur. Phys. J. C \textbf{76} (2016) 246.

\bibitem{zubair} M. Zubair, F. Kousar and S. Bahamonde, gr-qc/arXiv:1712.05699v1 (2017).

\bibitem{peter} P. K. F. Kuhfittig, Ind. J. of Phys. \textbf{92} (2018) 1207.

\bibitem{novi} I. D. Novikov, Physics-Uspekhi \textbf{61} (2018) 280 .


\bibitem{Zangeneh1}  	
M. K. Zangeneh, F. S. N. Lobo and N. Riazi, Phys. Rev. D \textbf{90} (2014) 024072.

\bibitem{Mehdizadeha}  	
M. R. Mehdizadeh,  M. K. Zangeneh and F. S. N. Lobo, Phys. Rev. D \textbf{91} (2015) 084004.

\bibitem{Zangeneh} 	
M. K. Zangeneh, F. S. N. Lobo and M. H. Dehghani, Phys. Rev. D \textbf{92} (2015) 124049.

\bibitem{Moraes}  	
P. H. R. S. Moraes, R. A. C. Correa and R. V. Lobato, JCAP \textbf{1707} (2017) 029.
\bibitem{Bejarano} 	
C. Bejarano, F. S. N. Lobo, G. J. Olmo and D. Rubiera-Garcia, Eur. Phys. J. C \textbf{77} (2017) 776.

\bibitem{Mehdizadeh}  	
M. R. Mehdizadeh and F. S. N. Lobo, Phys. Rev. D \textbf{93} (2016) 124014.




\bibitem{Rogatko}  	
M. Rogatko, Phys. Rev. D \textbf{97} (2018) 024001.

\bibitem{Paul}  	
B. C. Paul and A. S. Majumdar, Class. Quant. Grav. \textbf{35} (2018) 065001.


\bibitem{Shaikh} 
R. Shaikh, Phys. Rev. D \textbf{98} (2018) 024044.


\bibitem{Ovgun}  	
A. Ovgun, K. Jusufi and Izzet Sakalli, arXiv: 1804.09911[gr-qc] (2018)

\bibitem{Tsukamoto}  	
N. Tsukamoto and T. Kokubu, Phys. Rev. D \textbf{98} (2018) 044026.

\bibitem{barros} B. J. Barros and F. S. N. Lobo, Phys. Rev. D \textbf{98} (2018) 044012.	

\bibitem{cataldo} M. Cataldo, L. Liempi and P. Rodríguez, Eur. Phys. J. C. \textbf{77} (2017) 1.
\bibitem{noj1} S. Nojiri and S. D. Odintosov, Phys. Rev. D \textbf{68} (2003) 123512.

\bibitem{storiou6} T. P. Sotiriou, Phys. Rev. D. \textbf{73} (2006) 063515.

\bibitem{meng4} X. H. Meng and P. Weng, Phys. Lett. B. \textbf{584} (2004) 1.

\bibitem{hu} W. Hu and I. Sawicki, Phys. Rev. D \textbf{76} (2007) 064004.


\bibitem{Dutta}  	
K. Dutta, S. Panda and A. Patel, Phys. Rev. D \textbf{94} (2016) 024016.




\bibitem{Sharif}  	
M. Sharif and I. Nawazish,  Annals Phys. \textbf{389} (2018) 283.

\bibitem{cao} S. L. Cao, S. Li, H. R. Yu and T. J. Zhang, RAA, \textbf{18} (2018) 26
	\end{thebibliography}
\end{document}